\documentclass[a4paper,11pt]{article}
\pdfoutput=1 % if your are submitting a pdflatex (i.e. if you have
\usepackage{jheppub} % for details on the use of the package, please
\usepackage[T1]{fontenc} % if needed
\usepackage{braket}
\usepackage{csquotes}
\usepackage{nameref}
\usepackage{amsmath}
\usepackage{xspace}
\newcommand{\ee}{\mathrm{e}}
\newcommand{\ii}{\mathrm{i}}
\newcommand{\dd}{\mathrm{d}}
\newcommand{\tr}{\operatorname{tr}}
\newcommand{\lp}{l_{\mathrm{P}}}
\newcommand{\ls}{l_{*}}
\newcommand{\rhorizon}{r_{\mathrm{h}}}
\newcommand{\pd}{\partial}

\newcommand{\renyi}{R{\'e}nyi\xspace}
\newcommand{\Li}{\mathrm{Li}} % Polylogarithm function

\arxivnumber{2503.09070} % if you have one

\title{\boldmath Symmetry-Resolved Entanglement Entropy in Higher Dimensions}

\author[a]{Yuanzhu Huang}
\author[a]{and Yang Zhou}

\affiliation[a]{Department of Physics and Center for Field Theory and Particle Physics, Fudan University, \\Shanghai 200433, China}

\emailAdd{huangyz21@m.fudan.edu.cn}
\emailAdd{yang\_zhou@fudan.edu.cn}

\abstract{
  We present a method to compute the symmetry-resolved entanglement entropy of spherical regions in higher-dimensional conformal field theories. By employing Casini-Huerta-Myers mapping, we transform the entanglement problem into thermodynamic calculations in hyperbolic space. This method is demonstrated through computations in both free field theories and holographic field theories. For large hyperbolic space volume, our results reveal a universal expansion structure of symmetry-resolved entanglement entropy, with the equipartition property holding up to the constant order. Using asymptotic analysis techniques, we prove this expansion structure and the equipartition property in arbitrary dimensions.
  
}

\begin{document}
\maketitle
\flushbottom

\section{Introduction}\label{sec:intro}

Entropy measures the uncertainty associated with the state of a system~\cite{Shannon}. In quantum physics, entanglement can introduce such uncertainty. Consider a bipartite quantum system with a Hilbert space that can be expressed as the product of the subspaces of two subsystems \(\mathcal{H} = \mathcal{H}_A \otimes \mathcal{H}_B\). If the composite system is in a pure state while its individual subsystems, $A$ and $B$,  are in mixed states, then the state is classified as entangled. The von Neumann entropy, the quantum counterpart of classical Shannon entropy, is defined as
\begin{equation}
  S(\rho_A) = -\tr[\rho_A\log\rho_A],
\end{equation}
and it quantifies the uncertainty in the reduced mixed state of subsystem $A$. Since this uncertainty arises from the entanglement between $A$ and $B$, the von Neumann entropy also serves as a measure of entanglement and is thus referred to as entanglement entropy (EE).

Symmetry is another important feature of physical systems, making it natural to study entanglement through block-diagonalization in the presence of symmetry as follows.
Consider a bipartite system with global U(1) symmetry with conserved charge \(Q\), which is a sum of charge of the two subsystems, \(Q=Q_A+Q_B\). When the total system density matrix \(\rho\) commutes with the conserved charge \(Q\), it can be proved that, by partial trace, the subsystem density matrix still commute with the corresponding charge operator, i.e., \([\rho_A,Q_A] = [\rho_B,Q_B]=0\). Therefore, the subsystem density matrix  \(\rho_A\) will be block-diagonal in the eigenbases of the conserved charge \(Q_A\):
\begin{equation}
  \rho_A = \bigoplus_q \tilde{\rho}_A(q),
\end{equation}
where each \(\tilde{\rho}_A(q)\) is the projection of \(\rho_A\) onto the sector with charge \(q\), and the probability for the subsystem \(A\) occupying the \(q\) sector is given by \(p_A(q) = \tr[\tilde{\rho}_A(q)]\).
It is natural to study the entanglement entropy within each charge sector, called \emph{symmetry-resolved entanglement entropy} (SREE)~\cite{Goldstein:SRE}
\begin{equation}\label{eq:SREE_defn}
  S(q) = -\tr\left[\rho_A(q)\log \rho_A(q)\right],
\end{equation}
where \(\rho_A(q) = \frac{\tilde{\rho}_A(q)}{\tr[\tilde{\rho}_A(q)]}\) is the normalized block-diagonal density matrix.
For simplicity and consistency, in this paper, density matrices without tildes \enquote{\(\sim\)} will denote normalized density matrices with trace 1.
SREE is a measure of uncertainty in each sector and reflects how symmetry influences the structure of entanglement. The total von Neumann entropy can be resolved into two parts as follows:
\begin{equation}\label{eq:coarse_quantum}
  S(\rho_A) = -\sum_{q}p_A(q) \log p_A(q) + \sum_q p_A(q) S(\rho_A(q)),
\end{equation}
The first term in this decomposition formula is the Shannon entropy of the probability distribution \(\{p_A(q)\}\), and the second term is the average entanglement entropy of all sectors. For a more detailed explanation of symmetry resolution of entanglement and a perspective closer to Shannon's original idea~\cite{Shannon}, readers may refer to Appendix \ref{sec:SRE}.

To compute SREE, we employ the replica trick~\cite{Holzhey:geometric_entropy,Calabrese:EE_QFT}, a useful approach in entanglement entropy calculations. Using this method, we first compute the \renyi entropy, which is defined as
\begin{equation}
  S_n = \frac{1}{1-n} \log \tr[\rho_A^n],
\end{equation}
and then obtain the entanglement entropy by analytically continuing \(n\rightarrow 1\). This method is also helpful when computing SREE. We will first compute symmetry-resolved \renyi entropy (SRRE), which is defined as \renyi entropy within a charge sector,
\begin{equation}\label{eq:SRRE_defn}
  S_n(q) = \frac{1}{1-n} \log \tr[\rho_A(q)^n],
\end{equation}
and then take the limit \(n \rightarrow 1\) to obtain SREE.

The computation of SRRE \(S_n(q)\) requires evaluating \(\tr[\rho_A(q)^n]\), which is often a non-trivial task for generic systems. An approach was proposed in~\cite{Goldstein:SRE} to address this problem. This approach involves computing the following quantity called \enquote{charged moment}:
\begin{equation}
  \mathcal{Z}_n(\mu) = \tr\left[\rho_A^n \ee^{\ii \mu Q}\right].
\end{equation}
Performing an inverse Fourier transform will give us the trace in sectors, which we will call symmetry-resolved partition functions,
\begin{equation}
  \tr[\tilde{\rho}_A(q)^n] \equiv \mathcal{Z}_n(q) = \int_{-\pi}^{\pi} \frac{\dd \mu}{2\pi} \mathcal{Z}_n(\mu) \mathrm{e}^{-\ii \mu q}.
\end{equation}
Then SRRE can be computed as
\begin{equation}\label{eq:SRRE_from_Znq}
  S_n(q) = \frac{1}{1-n} \log \frac{\mathcal{Z}_n(q)}{\mathcal{Z}_1^n(q)}.
\end{equation}

In quantum field theory, \(\mathcal{Z}_n (\mu)\) is computed as a path integral,  with 2-dimensional conformal field theory (CFT) offering tractable examples~\cite{Goldstein:SRE}.
Some other results of charged moments and SREE includes computations in Wess-Zumino-Witten (WZW) models~\cite{Calabrese:SREE_in_WZW} and in excited states of CFT~\cite{Capizzi:SRE_of_excited}, and analysis for the charged moment in 2-dimensional Dirac and complex scalar fields~\cite{Murciano:SR_2DFreeQFT}, among others.
In the framework of AdS/CFT correspondence~\cite{Maldacena:AdSCFT,Witten:AdS_and_holography,GKP}, the holographic dual of the charged moment (initially proposed in~\cite{Belin:holo_charged_renyi} from a perspective distinct from SREE) has been explicitly linked to SREE in~\cite{Zhao:SRE_in_AdSCFT}. The proposal in~\cite{Zhao:SRE_in_AdSCFT} was later examined in~\cite{Weisenberger:SRE_excited_and_2interval}.
Some studies have extended symmetry resolution of entanglement beyond conventional group-like symmetries to categorical symmetries~\cite{Saura-Bastida:cat_SRE,Das:gener_SRE,Choi:noninvert_SR_ALC}.
Besides entanglement entropy, other entanglement measures were also discussed in the context of symmetry-resolved entanglement, such as the symmetry resolution of relative entropy~\cite{Chen:SR_rela,Capizzi:SR_rela}, the capacity of entanglement~\cite{Arias:capacity}, and mutual information~\cite{Parez:ExactQuench_SRE}. For other related studies on symmetry-resolved entanglement, see~\cite{Horvath:SRE_integrable,Zhao:W3_higher_spin, Baiguera:shape_deform_charged_Renyi,Ares:multi_chargedMoment,Murciano:SR_Page,DiGiulio:BCFT_to_SRE,Berthiere:reflEntrp_CrsNormNegat,Kusuki:SREE_BCFT,Rottoli:Ising_crossover,Pirmoradian:SRE_local_nonLocal,Feldman:SRE_lattice,Bianchi:nonAbel_SREE}.

One finding about SREE is that for the system with U(1) global symmetry, the expansion of SREE \(S(q)\) is independent of the value of the conserved charge \(q\) up to a certain order. This property is known as \emph{equipartition}~\cite{Xavier:Equipartition,Goldstein:SRE}.
For example, consider a massless compact boson, which is a conformal field theory with U(1) global symmetry and central charge \(c=1\). For a subsystem that is an interval of length \(L\), the SREE is~\cite{Xavier:Equipartition,Goldstein:SRE}
\begin{equation}\label{eq:SREE_2D}
  S(q) = S - \frac{1}{2} \log \left(\frac{2K}{\pi} \log L\right) -\frac{1}{2}+ \mathcal{O}(L^{-1}),
\end{equation}
where \(K\) is a constant related to the compactification radius of the boson. This formula is expanded in terms of the length \(L\) of the interval, and up to the constant order,  \(S(q)\) is independent of the conserved charge, i.e., it exhibits the property of equipartition.

Besides advancements in exploring symmetry-resolved entanglement measures in (1+1)-dimensional cases, several studies have also discussed higher-dimensional cases, such as Fermi gas in 2 spatial dimensions~\cite{Tan:particleNumFluct_2DFermiGas}, free non-relativistic massless fermions and free bosons in 2 spatial dimensions by dimensional reduction~\cite{Murciano:SRE_2D_dimReduct}, and free Fermi gas in general dimensions using Widom conjecture~\cite{Fraenkel:SRE_1D_beyond}.
A result for the charged moments across a hyperplane in \(d\) Euclidean dimensions has also been presented~\cite{Murciano:SR_2DFreeQFT}.
However, most studies explored cases in specific dimensions, and the universal structures of symmetry-resolved entanglement in general dimensions is not clear.
In this paper, we propose a method to compute SREE that can be applied in general higher dimensions.
The method mainly relies on the well-known technique of computing entanglement entropy for a spherical region in a CFT by conformal mapping~\cite{Myers:cthm_holo,CHM:towards,Hung:holo_renyi}.
The conformal mapping relates the domain of dependence of the spherical region to the hyperbolic cylinder \(\mathbb{R} \times H^{d-1}\).
To compute charged moment, we also employ the method in~\cite{Belin:holo_charged_renyi}, incorporating the new insertion \(\mathrm{e}^{\ii \mu Q}\) in the conformal mapping, and then the charged moment will be converted to a grand-canonical partition function on the hyperbolic space.
While calculating  this partition function is still difficult for generic CFT, we examine two types of theories where explicit computation is feasible.
The first type is free field theories, in which computation of partition function, and hence the charged moment, can be done exactly, and then SREE can be computed from the charged moments following steps as stated above.
The second type of theories in which the partition function can be computed is holographic theories. Following~\cite{Belin:holo_charged_renyi}\footnote{Strictly speaking, the quantities considered in~\cite{Belin:holo_charged_renyi} are not exactly the same as symmetry-resolved entanglement measures. However, the techniques can still apply here.}, the grand-canonical partition function will be equivalent to the grand-canonical partition function of a charged black hole due to \(\mathrm{e}^{\ii \mu Q}\) in the charged moment.
Due to the universal applicability of the conformal mapping, our method can be applied to general dimensions.

The main computational results of this paper reveal that the SREE exhibits an expansion structure of the following form:
\begin{equation}
  S(q) = S - \# \log V + \mathcal{O}\left(V^0\right) + q\text{-dependent terms},
\end{equation}
where \(V\) is the volume of hyperbolic space in our computation method.
In this expansion, the leading term is the unresolved entropy, which is proportional to \(V\). Both the subsequent logarithmic term and the constant term remain independent of the charge \(q\), showing that equipartition holds up to constant order. The charge dependence appears in higher-order terms that vanish in the large \(V\) limit.

The paper is organized as follows:
In Section~\ref{sec:methods}, we introduce the method of computing SREE of a spherical region in higher-dimensional CFTs.
In Section~\ref{sec:free_calc}, we apply this method to free field theory to compute SREE.
In Section~\ref{sec:holo_calc}, we apply this method to holographic CFT in higher dimensions to compute SREE.
The results in Section~\ref{sec:free_calc} and~\ref{sec:holo_calc} give the same universal asymptotic structure of charged moments, SRRE, and SREE. In particular, the property of equipartition of SREE at leading orders is again discovered.
In Section~\ref{sec:structure}, we summarize and analyze the universal expansion structure of SRRE and SREE, and show that the structure is a result of choosing the size of the subsystem as the expansion parameter, and details of the underlying field theory is irrelevant.
We conclude with a discussion in Section~\ref{sec:discussion}.
In Appendix \ref{sec:SRE}, we motivate symmetry-resolved entanglement from the perspective of entropy coarse-graining, following the primitive thoughts of Shannon.
In Appendix \ref{sec:higher_asymp}, we provide the explanation and derivation of an asymptotic expansion formula used in explicit computations.

\section{Methods}\label{sec:methods}
In this section, we present the method for computing symmetry-resolved entanglement entropy of spherical regions in higher-dimensional CFT. Following~\cite{CHM:towards,Hung:holo_renyi,Belin:holo_charged_renyi}, we describe the conformal mapping that maps the charged moment to the grand-canonical partition function in hyperbolic space. The computation of this thermal partition function can be done in several cases, which will be shown in Section \ref{sec:free_calc} and Section \ref{sec:holo_calc}.

Consider a \(d\)-dimensional CFT in flat space in its vacuum state, and choose a spherical region \(A\) of radius \(R\) as a subsystem. Via the conformal mapping presented in~\cite{CHM:towards}, the domain of dependence \(\mathcal{D}\) of the spherical region \(A\) can be mapped to a hyperbolic cylinder \(\mathbb{R} \times H^{d-1}\), where \(\mathbb{R}\) represents the time direction and \(H^{d-1}\) represents \((d-1)\)-dimensional hyperbolic space.
Furthermore, the reduced density matrix of the spherical region is mapped to the thermal state density matrix in the hyperbolic space, with temperature being \(T_0 = \beta_0^{-1}= \frac{1}{2\pi R}\):
\begin{equation}\label{eq:rho_to_thermal}
  \rho_{A} = \frac{1}{Z(T_0)} U^{-1} \mathrm{e}^{-\beta_0 H} U,
\end{equation}
where \(Z(T_0) = \tr\left[\mathrm{e}^{-\beta_0 H}\right]\) is the thermal partition function, and \(U\) is the unitary operator implementing the conformal transformation.

Via this conformal mapping, the entanglement-related quantities of the original spherical region \(A\) transforms into the thermodynamic quantities in hyperbolic space. For free field theory, the path integral in hyperbolic space can be solved exactly to calculate the corresponding entanglement entropy. For holographic field theory, the thermodynamic entropy in hyperbolic space can be further expressed as the black hole entropy in the dual gravitational theory through holographic duality.

This method extends naturally to the computation of \(\tr[\rho^n_A]\). The key idea is that raising the density matrix to a power is equivalent to changing the corresponding temperature of the thermal density matrix~\cite{Hung:holo_renyi}:
\begin{equation}
  \rho^n_{A} = \frac{1}{Z(T_0)^n}U^{-1} \mathrm{e}^{-n \beta_0 H} U.
\end{equation}

Furthermore, in the presence of global symmetry, the conformal transformation also acts on the charge operator \(Q_A\):
\begin{equation}
  Q_A = U^{-1} Q U,
\end{equation}
where \(Q\) represent the charge operator in hyperbolic space \(H^{d-1}\).
Based on the conformal mapping relations of the density matrix \(\rho_A\) and conserved charge \(Q_A\), the charged moments for the spherical region \(A\), defined as \( \mathcal{Z}_n(\mu) = \tr[\rho_{A}^n \mathrm{e}^{\ii \mu Q_A}] \), can be expressed in terms of thermodynamic quantities in hyperbolic space as
\begin{equation}
  \mathcal{Z}_n(\mu) = \frac{1}{Z(T_0)^n} \tr\left[\mathrm{e}^{-n\beta_0 H} \mathrm{e}^{\ii \mu Q}\right].
\end{equation}

We then define the following grand-canonical partition function:
\begin{equation}\label{eq:grandZ}
  Z\left(n\beta_0, \frac{\ii \mu}{n\beta_0}\right) = \tr\left[\mathrm{e}^{-n\beta_0 H + \ii  \mu Q}\right],
\end{equation}
where the corresponding inverse temperature is
\begin{equation}\label{eq:T_n_relation}
  \beta = n\beta_0 = 2\pi n R.
\end{equation}
Here, the notation for \(Z\) aligns with that of a grand-canonical partition function in thermodynamics, where we have \(Z(\beta,\Phi) = \tr[\mathrm{e}^{-\beta(H- \Phi Q)}]\).
The charged moments can thus be expressed in terms of the grand-canonical partition function as
\begin{equation}\label{eq:mcalZ_from_Z}
  \mathcal{Z}_n(\mu) = \frac{Z\left(n\beta_0, \frac{\ii \mu}{n\beta_0}\right)}{Z(\beta_0,0)^n}.
\end{equation}

The computation of charged moments requires \emph{imaginary} chemical potential in the grand-canonical partition function.
As in~\cite{Belin:holo_charged_renyi}, the insertion of \(\mathrm{e}^{\ii \mu Q}\) is equivalent to a gauge field with integral around the Euclidean time circle\footnote{In the literature, \(\mu\) here is sometimes also referred to as a "chemical potential", although here it is dimensionless and does not have the same dimensionality as a chemical potential in thermodynamics, where it carries units of energy.} \(\oint B_{\tau}\dd\tau = \mu\), and the effect of this gauge field is to modify the boundary condition, so that when a particle traverses the Euclidean time circle, it will acquire an extra phase \(\mathrm{e}^{\ii \oint B} = \mathrm{e}^{\ii \mu}\)~\cite{Roberge:gauge_Im,Belin:holo_charged_renyi}.

We also note that there is a simplification in the computation. As shown in Equation~\eqref{eq:SRRE_from_Znq}, the calculation of SRRE requires only the ratio \(\frac{\mathcal{Z}_n(q)}{\mathcal{Z}_1^n(q)}\) rather than individual values of \(\mathcal{Z}_n(q)\). It can be easily shown that, since the normalization factor \(Z(\beta_0,0)^n\) is a \(\mu\)-independent number raised to the power of \(n\), it will cancel out in the ratio.
Therefore, we can use unnormalized\footnote{In this language, \(\mathcal{Z}_n(\mu)\) is called \enquote{normalized} charged moment, since \(\mathcal{Z}_1(0)=1\).} charged moments, defined as \(\tilde{\mathcal{Z}}_n(\mu) := Z\left(n\beta_0, \frac{\ii \mu}{n\beta_0}\right)\), and correspondingly define the unnormalized symmetry-resolved partition function as
\begin{equation}\label{eq:tildeZnq}
  \tilde{\mathcal{Z}}_n(q) := \int_{-\pi}^{\pi} \frac{\dd \mu}{2\pi} \tilde{\mathcal{Z}}_n(\mu) \ee^{-\ii \mu q} =  \int_{-\pi}^{\pi} \frac{\dd \mu}{2\pi} Z\left(n\beta_0, \frac{\ii \mu}{n\beta_0}\right) \ee^{-\ii \mu q}.
\end{equation}
Thus, the symmetry-resolved \renyi entropy can be expressed as
\begin{equation}\label{eq:SRRE_from_tildeZnq}
  S_n(q) = \frac{1}{1-n} \log \frac{\tilde{\mathcal{Z}}_n(q)}{\tilde{\mathcal{Z}}_1^n(q)}.
\end{equation}

Let's now summarize some results of the hyperbolic space volume and describe the expansion limit we will use in explicit computations. In the computation method presented above, the hyperbolic space volume \(V\) appears in the expression for the charged moment \(\mathcal{Z}_n(\mu)\), as we will see later in specific computational examples.
Furthermore, the volume \(V\) will appear in symmetry-resolved entanglement entropy. The hyperbolic space has infinite volume, as can be seen from the following metric~\cite{Hung:holo_renyi}:
\begin{equation}
  \dd \Sigma^2_{d-1} = \frac{\dd y^2}{y^2-1}+(y^2-1)\dd \Omega_{d-2}^2,
\end{equation}
where \(\dd\Omega_{d-2}^2\) is the line element of the unit sphere \(S^{d-2}\), and the coordinate range is \(y \in (1,+\infty)\).
To render physical quantities finite, regularization is required. In the cutoff regularization, the upper limit of the \(y\) integral is set to a finite value \(Y\). In the original flat space, ultraviolet (UV) divergence occurs at the boundary of the spherical region due to the sharp cut of the subsystem. To address this divergence, a short-distance cutoff \(\delta\) is introduced, and contributions are considered down to a radius \(r = R - \delta\), where \(\delta/R \ll 1\), rather than extending to the boundary of the spherical region at \(r = R\). Consistency requires that the two cutoffs be related by the conformal mapping between the two spaces and finally gives~\cite{CHM:towards,Hung:holo_renyi}
\begin{equation}
  Y = \frac{R}{\delta}
\end{equation}

With a finite integral limit \(Y\), the volume of hyperbolic space is regularized as
\begin{equation}\label{eq:vol_hyper}
  \begin{aligned}
    V & = \Omega_{d-2} \int_{1}^{Y}(y^2-1)^{(d-3)/2}\dd y                                                                                       \\
      & = \frac{\Omega_{d-2}}{d-2}\left[ \frac{R^{d-2}}{\delta^{d-2}} - \frac{(d-2)(d-3)}{2(d-4)} \frac{R^{d-4}}{\delta^{d-4}} +\cdots \right],
  \end{aligned}
\end{equation}
where \(\Omega_{d-2} = \frac{2\pi^{(d-1)/2}}{\Gamma((d-1)/2)}\) represents the area of unit sphere \(S^{d-2}\).
When computing entanglement entropy, a standard case often considered is the limit where the subsystem size \(R\) is much larger than the cutoff \(\delta\), i.e., \(R\gg \delta\), corresponding to a continuum limit. This is the usually adopted limit in field-theoretic calculations of entanglement entropy, and is also how theoretical predictions can be compared with numerical results (See, for example,~\cite{Vidal:entanglement}).
From the above expansion \eqref{eq:vol_hyper}, we see that it naturally corresponds to a large \(V\) limit.
In the conformal mapping we use in this paper, the state of the spherical region is mapped to a thermal state. From the point of view of the thermal state on hyperbolic space, the large \(V\) limit can also be seen as the thermodynamic limit.
Since the transform in Equation~\eqref{eq:tildeZnq} generally lacks an analytical solution, in computations later on, we will employ asymptotic analysis, using \(V\) as the large expansion parameter.

Lastly, the universal term, which is independent of cutoff choosing or regularization method, is~\cite{CHM:towards}
\begin{equation}\label{eq:V_univ}
  V_{\text{univ}} = \frac{\pi^{d/2}}{\Gamma(d/2)} \times
  \begin{cases}
    (-)^{\frac{d}{2}-1} \frac{2}{\pi} \log (2 R / \delta), & d \, \text{even}, \\
    (-)^{\frac{d-1}{2}},                                   & d \, \text{odd}.
  \end{cases}
\end{equation}
The divergence structure and universal terms of entanglement entropy in the spherical region are encoded in \(V_{\mathrm{univ}}\) \cite{CHM:towards}.

\section{Free Field Computation}\label{sec:free_calc}
We apply the conformal mapping method discussed in Section~\ref{sec:methods} to calculate the charged moment of a spherical region in free CFT, and we derive symmetry-resolved entanglement entropy from the charged moment.

As mentioned in the last section, computing the charged moment \(\mathcal{Z}_n(\mu) = \tr[\rho^n \mathrm{e}^{\ii \mu Q}]\) is equivalent to computing the partition function with imaginary chemical potential. For free field computations in this section, the partition function and the charged moment can be expressed in terms of the heat kernel~\cite{Vassilevich:heatkernel_manual}. Here we outline the ingredients needed for these computations. A pedagogical introduction to the heat kernel, along with related results, is provided in Appendix \ref{sec:heat_kernel_essentials}.

In the path integral approach, this imaginary chemical potential can be seen as modifying the boundary condition of the field~\cite{Belin:holo_charged_renyi,Roberge:gauge_Im}. For example, a scalar boson traversing the Euclidean time circle will acquire an extra phase~\cite{Roberge:gauge_Im,Belin:holo_charged_renyi,Goldstein:SRE}:
\begin{equation}\label{eq:twisted_BC}
  \phi(\tau+\beta) = \mathrm{e}^{\ii \mu} \phi(\tau).
\end{equation}
The modification to the boundary condition is captured effectively by the heat kernel, which also yields a phase \(\mathrm{e}^{\ii \mu}\) around the Euclidean time direction~\cite{Belin:holo_charged_renyi}:
\begin{equation}\label{eq:K_period_mu}
  K_{S^1_\beta \times H^{d-1}}(x+\beta,y,t) = \mathrm{e}^{\ii \mu} K_{S^1_\beta \times H^{d-1}}(x,y,t),
\end{equation}
where \(x\) denotes the coordinate along the periodic Euclidean time direction \(S_{\beta}^1\) (the subscript \(\beta\) indicates the periodicity), and \(y\) represents the coordinates on \(H^{d-1}\).
Note that \(t\) in the above formula represents a parameter in the heat kernel, rather than time.
For convenience, we now set the hyperbolic space radius to be dimensionless \(R=1\). Consequently, the temperature relevant to \(\mathcal{Z}_n(\mu) = \tr[\rho^n \mathrm{e}^{\ii \mu Q}]\) is \(T = \frac{1}{2\pi n}\).

On the product manifold \(S^1_\beta \times H^{d-1}\), the heat kernel of a free scalar field takes the form of a product of the heat kernels on the submanifolds:
\begin{equation}
  K_{S^1_\beta \times H^{d-1}}(\{x_1, x_2\},\{y_1,y_2\},t) = K_{S^{1}_{\beta}}(x_1,y_1,t)\, K_{H^{d-1}}(x_2,y_2,t),
\end{equation}
where the factor on \(S^1_{\beta}\) (which should satisfy the twisted boundary conditions \eqref{eq:K_period_mu}) is given by~\cite{Grigor'yan:heatKer_hyperbolic,Vassilevich:heatkernel_manual,Belin:holo_charged_renyi} (see Appendix \ref{sec:heat_kernel_essentials} for a derivation)
\begin{equation}\label{eq:K_S1_mu}
  K_{S^1_{\beta}}(x_1,x_2,t) = \frac{1}{\sqrt{4\pi t}} \sum_{k \in \mathbb{Z}} \exp\left[-\frac{(x_1 - x_2 - 2\pi n k)^2}{4t} + \ii k \mu \right],
\end{equation}
and the specific form of the heat kernel on hyperbolic space \(K_{H^{d-1}}(x,y,t)\) is determined by the theory under consideration.

Following the normalization discussion in Section \ref{sec:methods}, we calculate SREE using the unnormalized charged moment \(\tilde{\mathcal{Z}}_n(\mu) = \mathrm{e}^{-F_{n}(\mu)}\). For complex scalar theories considered in the following subsections, the free energy \(F_n(\mu)\) can be expressed in terms of the heat kernel as (for more details, see the following subsections)
\begin{equation}
  F_n(\mu) = - \int_{0}^{+\infty} \frac{\dd t}{t} \int \dd x \,  \dd^{d-1}y \, K_{{S^1_\beta \times H^{d-1}}}(\{x,y\},\{x,y\},t).
\end{equation}
We now consider specific theories to determine the explicit form of \(K_{S^{1}_{\beta}\times H^{d-1}}\), which will allow us to compute the free energy and charged moment, and ultimately obtain the SREE.

\subsection{4D scalar field}\label{sec:free_calc_4DScalar}
We consider a conformally coupled complex scalar field in \(d=4\) dimensions. The Euclidean action is given by
\begin{equation}
  S = \int \dd^4x \sqrt{g} \left[ |\nabla\phi|^2 + \frac{1}{6} \mathcal{R} |\phi|^2 \right],
\end{equation}
where \(\mathcal{R}\) is the Ricci scalar curvature, and its coupling with \(|\phi|^2\) with a coefficient \(\frac{1}{6}\) ensures the conformal symmetry in 4D.
We consider a complex scalar field rather than a real scalar field, since the former admits a global U(1) symmetry \(\phi\rightarrow \phi\,\ee^{\ii\alpha}\) but the (non-compact) real scalar does not.
The kinetic differential operator in this theory is the Laplace operator supplemented by a conformal coupling term. On an odd-dimensional hyperbolic space \(H^{d-1}\), the corresponding heat kernel is~\cite{Grigor'yan:heatKer_hyperbolic}
\begin{equation}
  \begin{aligned}
    K_{H^{d-1}}(y_1,y_2,t) & = \frac{1}{(4\pi t)^{\frac{d-1}{2}}} \frac{\rho(y_1,y_2)}{\sinh \rho (y_1,y_2)} \mathrm{e}^{- \frac{\rho^2(y_1,y_2)}{4t}},
  \end{aligned}
\end{equation}
where \(\rho(y_1,y_2)\) denotes the geodesic distance between points \(y_1\) and \(y_2\). When the two points coincide, \(\rho(y,y)=0\), yielding
\begin{equation}
  K_{H^{d-1}} (y,y,t) = \frac{1}{(4\pi t)^{\frac{d-1}{2}}}.
\end{equation}
The heat kernel on \(S_{\beta}^{1}\) is given by Equation \eqref{eq:K_S1_mu}, and we provide a derivation of it in Appendix~\ref{sec:heat_kernel_essentials}.

The free energy of the complex scalar field on \(S_{\beta}^1 \times H^{3}\) is given by
\begin{equation}
  \begin{aligned}
    F_n(\mu) & = - \int_{0}^{+\infty} \frac{\dd t}{t} \int_{0}^{2\pi n} \dd x \int \dd ^{3}y \, K_{{S^1_\beta \times H^{3}}}(\{x,y\},\{x,y\},t) \\
             & \sim - \frac{V}{8 n^3 \pi ^5} \sum_{k \in \mathbb{Z}, k\ne 0} \frac{\mathrm{e}^{\ii k \mu}}{k^4}                                              \\
             & = - \frac{V}{8 n^3 \pi ^5}\left(\Li_4(\mathrm{e}^{\ii \mu}) + \Li_4(\mathrm{e}^{-\ii \mu})\right).
  \end{aligned}
\end{equation}
Several points about this expression need explanation.
Firstly, the \(k=0\) term in the sum produces a divergence proportional to \(n\) and independent of \(\mu\), and hence it will be eventually cancelled in the ratio \(\frac{\mathcal{Z}_n(q)}{\mathcal{Z}_1^n(q)}\) when computing SRRE. Therefore, the \(k=0\) term is omitted in the second line, and the remaining terms sum to polylogarithm functions \(\Li_n(z)\).
Secondly, the hyperbolic space volume \(V\) is also divergent; we formally retain it as the symbol \(V\), which can later be assigned a finite value using cutoff regularization.
Finally, the free energy \(F_n(\mu)\) is periodic, and within the period \(\mu \in [0,2\pi]\), the sum of the two polylogarithm functions simplifies as
\begin{equation}
  \Li_4(\mathrm{e}^{\ii \mu}) + \Li_4(\mathrm{e}^{-\ii \mu})  = \frac{1}{360} \left(8\pi ^4 - 60 \pi ^2 \mu ^2 + 60 \pi \mu^3 -15 \mu^4 \right),
\end{equation}
yielding
\begin{equation}
  F_n(\mu) = -\frac{\left(-15 \mu ^4+60 \pi  \mu ^3-60 \pi ^2 \mu ^2+8 \pi ^4\right) V}{2880 \pi ^5 n^3}, \quad \mu \in [0,2\pi].
\end{equation}
Values outside this interval are obtained by periodic translation.
\begin{figure}[htb]
  \centering
  \includegraphics[width=0.8\textwidth]{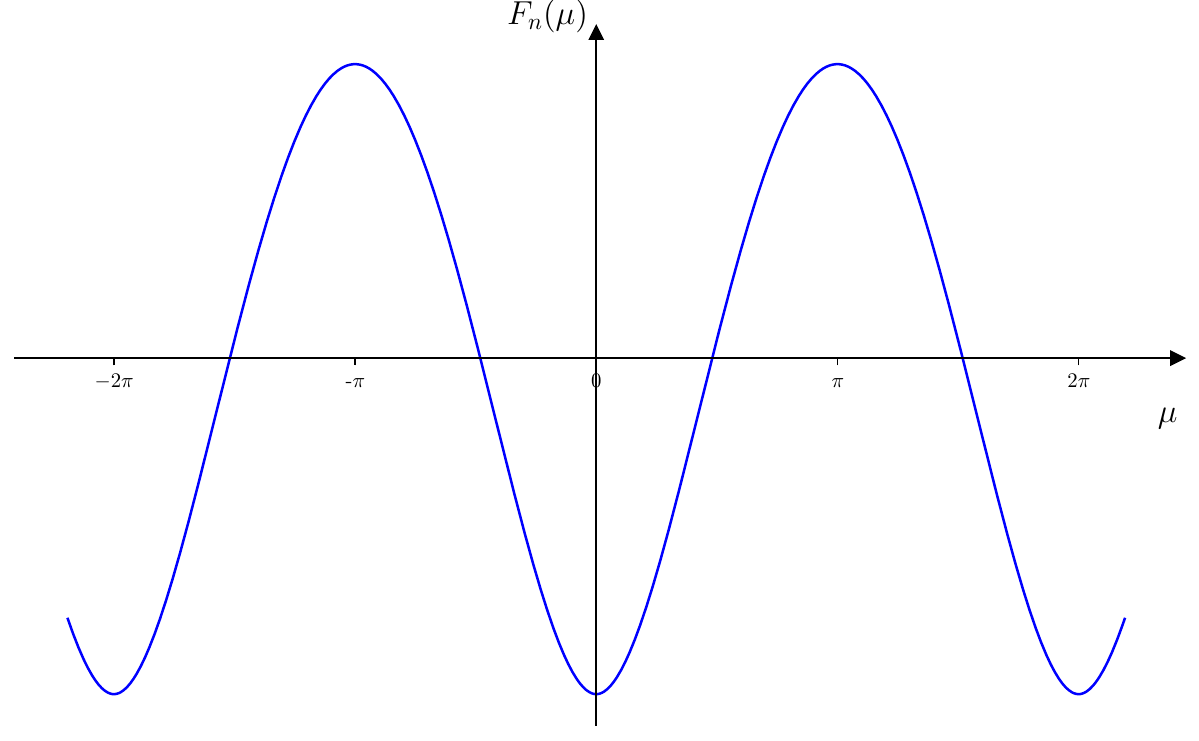}
  \caption{Plot of \(F_n(\mu)\) for complex free scalar field in 4D for \(n=1\). The qualitative shape and the minimum point (\(\mu=0\)) remain unchanged for different values of \(n\).}
  \label{fig:Fnmu_Scalar4D}
\end{figure}

As evident from Figure \ref{fig:Fnmu_Scalar4D}, \(F_n(\mu)\) exhibits a minimum at \(\mu=0\), corresponding to a maximum of \(\tilde{\mathcal{Z}}_n(\mu)\). This behavior is expected and can be understood through the following mathematical argument: The normalized charged moment \(\mathcal{Z}_n(\mu=0) = \tr\left[\mathrm{e}^{-\beta H}\right]\) represents a sum of positive terms \(\mathrm{e}^{-\beta E_i}\), which are aligned on the complex plane. In contrast, \(\mathcal{Z}_n(\mu) = \tr\left[\mathrm{e}^{-\beta H +\ii \mu Q}\right]\) involves summing these terms multiplied by different phase factors. This distribution leads to cancellations among terms, resulting in a sum with a smaller modulus compared to the sum at \(\mu=0\), i.e., \(\left|\tilde{\mathcal{Z}}_n(\mu)/\tilde{\mathcal{Z}}_n(0)\right| = \left|\mathcal{Z}_n(\mu)/\mathcal{Z}_n(0)\right|<1\).

Next, we need to perform the inverse Fourier transform~\eqref{eq:tildeZnq} from \(\mathcal{Z}_n(\mu)\) to \(\mathcal{Z}_n(q)\). Since the transform generally lacks an analytical solution, we employ asymptotic analysis, using the volume of hyperbolic space \(V\) as the expansion parameter.
Performing asymptotic analysis (the details of this technique is explained in Appendix \ref{sec:higher_asymp}), we obtain the unnormalized symmetry-resolved partition function to the subleading order
\begin{equation}\label{eq:tildeZnq_4DScalar}
  \tilde{\mathcal{Z}}_n(q) = 2 \sqrt{3} \pi  \exp\left(\frac{V}{360 \pi  n^3}\right) \frac{\sqrt{n^3}}{\sqrt{V}}  \left(1 - \frac{12 \pi  n^3 \left(\pi ^2 q^2-3\right)}{V} + \mathcal{O}\left(V^{-2}\right)\right).
\end{equation}
Finally, the asymptotic expansion of the SREE for a conformally coupled free complex scalar field in 4 dimensions in powers of \(V\) is given by
\begin{multline}
  S_n(q) = \frac{\left(n^3+n^2+n+1\right) V}{360 \pi  n^3}-\frac{1}{2}\log V - \frac{3\log n}{2(n-1)} + \log \left(2 \sqrt{3} \pi \right)  \\
  + \frac{12 \pi  n (n+1) \left(\pi ^2 q^2-3\right)}{V} + \mathcal{O}\left(V^{-2}\right).
\end{multline}
Taking the limit \(n\rightarrow 1\) yields the SREE
\begin{equation}\label{eq:SREE_4DScalar}
  S(q) = \frac{V}{90 \pi} - \frac{1}{2}\log V + \log \left(2 \sqrt{3}\pi \right) -\frac{3}{2}  + \frac{24 \pi  \left(\pi ^2 q^2-3\right)}{V} +\mathcal{O}(V^{-2}).
\end{equation}

Let us explain the terms in the above two equations. First, we note that the leading term in the SRRE is the unresolved \renyi entropy of a spherical region in \((3+1)\)-D CFT (where \(V\) is an expansion of \(R/\delta\)),
\begin{equation}
  S_n = \frac{\left(n^3+n^2+n+1\right) V}{360 \pi  n^3},
\end{equation}
and similarly, the leading term in SREE \(S(q)\) is the unresolved entanglement entropy
\begin{equation}
  S = \frac{V}{90 \pi}.
\end{equation}
This relationship can be interpreted as the equivalence of ensembles in the thermodynamic limit. While we are examining entanglement in a quantum system, the conformal mapping enables us to analyze it from a thermodynamic perspective.
From this perspective, \(S\) represents the total entropy of the system in an ensemble permitting charge fluctuation, whereas \(S(q)\) is in a \enquote{micro-canonical} ensemble in which charge takes a certain value. In the thermodynamic limit, which corresponds to the large subsystem limit, charge fluctuation becomes negligible, and \(S\) and \(S(q)\) agree at the leading order.
In formula, it follows from the definition of symmetry-resolved \renyi entropy in Equation \eqref{eq:SRRE_defn}:
\begin{equation}
  \begin{aligned}
    S_n(q) & = \frac{1}{1-n} \log \left( \frac{\mathcal{Z}_n}{\mathcal{Z}_1^n} \frac{\mathcal{Z}_n(q)}{\mathcal{Z}_n} \frac{\mathcal{Z}_1^n}{\mathcal{Z}_1^n(q)}\right) \\
           & = S_n + \frac{1}{1-n} \log \left(\frac{\mathcal{Z}_1^n(q)/\mathcal{Z}_1^n}{\mathcal{Z}_n(q)/\mathcal{Z}_n}\right).
  \end{aligned}
\end{equation}
Taking the limit \(n\rightarrow 1\) yields a similar relation for the symmetry-resolved entanglement entropy\footnote{Note the distinction between this equation and Equation \eqref{eq:coarse_quantum} from earlier. Here, \(S(q)\) means symmetry-resolved entanglement entropy for a specific \(q\), while Equation \eqref{eq:coarse_quantum} describes the relationship between \(S\) and the average of \(S(q)\).}:
\begin{equation}\label{eq:Sq_is_S_plus}
  S(q) = S + \lim_{n\rightarrow 1} \frac{1}{1-n} \log \left(\frac{\mathcal{Z}_1^n(q)/\mathcal{Z}_1^n}{\mathcal{Z}_n(q)/\mathcal{Z}_n}\right).
\end{equation}
Therefore, like the conventional entropy, the leading term of the SREE is proportional to \(V\), with its expansion leading term representing the area law.

The subsequent terms represent the difference between symmetry-resolved entanglement entropy and conventional entanglement entropy, corresponding to the second term in Equation \eqref{eq:Sq_is_S_plus}. Among these terms, the highest-order divergence is the logarithmic term \(-\frac{1}{2} \log V\). The coefficient \(-\frac{1}{2}\) comes from the asymptotic behavior of \(F_n(\mu)\) at \(\mu=0\). More specifically, the first derivative of \(F_n(\mu)\) vanishes at \(\mu=0\), while the second derivative does not. This causes \(\sqrt{V}\) in the denominator in Equation \eqref{eq:tildeZnq_4DScalar} and finally \(-\frac{1}{2} \log V\) in \(S_n(q)\) and \(S(q)\).

Motivated by the  universal terms in entanglement entropy, we explore whether analogous universal contributions might arise in SREE.
The logarithmic term \(-\frac{1}{45}\log(R/\delta)\) in original entanglement entropy is inherently a universal term, whose coefficient arises from type A anomaly~\cite{Solodukhin:EE_Conf_ExtGeom,Fursaev:SquashCone}; and the leading order of \(\log V\) also contributes a logarithmic term \(-\frac{1}{2}\log \left(\frac{R^{d-2}}{\delta^{d-2}}\right) = -\log (R/\delta)\), in which \(d=4\) in this case.
These two logarithmic terms combine to give a term \(S_{\log}(q) = -\frac{46}{45} \log\left(\frac{R}{\delta}\right)\), which is unaffected when changing \(\delta\).
Despite this robust mathematical structure, the physical significance of this combination of type A anomaly and dimensions remains unclear.
Further work is needed to determine whether they encode meaningful universal data or reflect subtleties inherent to the SREE construction itself.

We can further analyze the \(q\)-dependence of \(S(q)\). The expression \eqref{eq:SREE_4DScalar} shows that \(S(q)\) remains independent of the conserved charge \(q\) up to the constant term. This reproduces the equipartition property. To explicitly show the \(q\)-dependence, we expand the expression up to \(\mathcal{O}(V^{-1})\). This term depends on \(q\) through a \(q^2\) dependence, which is reasonable given that the complex scalar field is symmetric under charge conjugation.
Furthermore, this term sets a constraint on the validity of the large \(V\) approximation, which gives \(q\ll \frac{R}{\delta}\) in 4D scalar  case. However, the physical interpretation of this limitation is still unclear to us and requires investigation.

In summary, for a conformal free complex scalar field in 4D, the expansion of symmetry-resolved entanglement entropy \(S(q)\) in terms of \(V\) consists of the conventional entanglement entropy \(S\), plus the subleading term \(-\frac{1}{2} \log V\), followed by further subleading terms. The equipartition property holds up to the \(\mathcal{O}(V)\) term, and breaks down at the next higher-order term. Notably, the expansion structure of SREE aligns with the 2D compact scalar field results shown in Equation \eqref{eq:SREE_2D}. In particular, the equipartition properties holds up to the constant order. In Section \ref{sec:structure}, we will show that the expansion structure is a consequence of the expansion in \(V\), and that the dependence on \(q\) can only appear in terms beyond the constant order.

\subsection{2D scalar field}\label{sec:free_calc_2DScalar}
The method presented in Section~\ref{sec:methods} can also be applied in (1+1)-dimensional case. Consider a conformally coupled free non-compact scalar field in 2D. The \enquote{spherical region} in this dimension is an interval of length \(L = 2R\), and the boundary of it consists of two points. The hyperbolic space obtained by conformal mapping is \(H^1\).
The heat kernel on \(H^1\) is
\begin{equation}
  K_{H^1}(y,y,t) = \frac{1}{\sqrt{4\pi t}}.
\end{equation}
The free energy of the scalar field on \(S_{\beta}^1 \times H^1\) is
\begin{equation}
  \begin{aligned}
    F_n(\mu) & = - \int_{0}^{+\infty} \frac{\dd t}{t} \int_{0}^{2\pi n} \dd x \int \dd y\, K_{S_{\beta}^1 \times H^1}\left(\{x,y\},\{x,y\},t\right) \\
             & \sim -\frac{V }{2 \pi ^2 n} \left(\text{Li}_2\left(\ee^{-\ii \mu }\right)+\text{Li}_2\left(\ee^{\ii \mu }\right)\right).
  \end{aligned}
\end{equation}
As in 4D case, the \(k=0\) term in the sum produces a divergence but is omitted because it does not contribute to SREE. The sum of the two polylogarithm functions again simplifies to a polynomial, which in the interval \(\mu \in [0,2\pi]\) gives
\begin{equation}
  F_n(\mu) = -\frac{V}{12 \pi ^2 n} \left(3 \mu ^2-6 \pi  \mu +2 \pi ^2\right).
\end{equation}
Values outside this interval are obtained by periodic translation.
\begin{figure}[htb]
  \centering
  \includegraphics[width=0.8\textwidth]{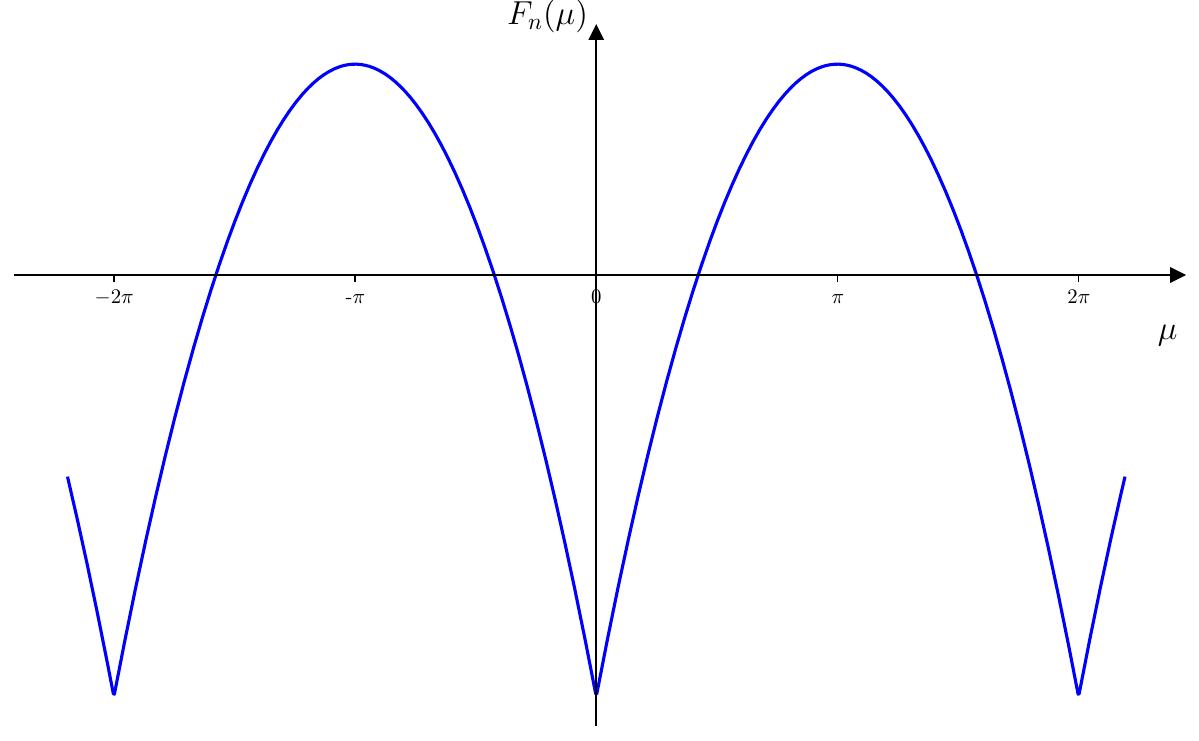}
  \caption{Plot of \(F_n(\mu)\) for complex free scalar field in 2D for \(n=1\). The qualitative shape and the minimum point (\(\mu=0\)) remain unchanged for different values of \(n\).}
  \label{fig:Fnmu_Scalar2D}
\end{figure}

As evident in Figure \ref{fig:Fnmu_Scalar2D}, \(\mu=0\) is a minimum point of \(F_n(\mu)\) and hence a maximum point of \(\tilde{\mathcal{Z}}_n(\mu)\), and is thus consistent with the fact that \(\left|\tilde{\mathcal{Z}}_n(\mu)/\tilde{\mathcal{Z}}_n(0)\right| < 1\) for \(\mu \ne 2 k \pi\, (k\in \mathbb{Z})\), as analyzed in 4D case.

Via inverse Fourier transform, in the large \(V\) expansion, the unnormalized symmetry-resolved partition function is
\begin{equation}
  \tilde{\mathcal{Z}}_n(q) = \frac{2n\, \exp\left(\frac{V}{6n}\right)}{V} \left(1 + \frac{2n}{V} + \frac{4n^2(-\pi^2 q^2+3)}{V^2} + \mathcal{O}(V^{-3})\right).
\end{equation}
Note that the denominator contains \(V\) rather than \(\sqrt{V}\) in 4D case. This ultimately results from \(F'(0)\ne 0\) and can be seen from the asymptotic analysis. A detailed explanation of this is given in Section~\ref{sec:structure_proof}.

The SRRE is given by
\begin{equation}\label{eq:SRRE_2DScalar}
  S_n(q) = \frac{1}{6}\left(1+\frac{1}{n}\right) V - \log V + \log 2 - \frac{\log n}{n-1} + \frac{2 n \left(2 \pi ^2 q^2-5\right)}{V^2} + \mathcal{O}(V^{-3})
\end{equation}
Taking the limit \(n\rightarrow 1\) yields the SREE
\begin{equation}\label{eq:SREE_2DScalar}
  S(q) = \frac{V}{3} - \log V + \log 2 - 1 + \frac{2\left(2 \pi ^2 q^2-5\right)}{V^2} + \mathcal{O}(V^{-3})
\end{equation}

To compare these two symmetry-resolved entropies with unresolved entropies, we make a note on the expression of \(V\) in the 2D case. Although the hyperbolic space \(H^1\) has the exactly same metric as the flat space \(\mathbb{R}^1\), in its parameterization, it has a logarithmic divergence term in its volume~\cite{Hung:holo_renyi}
\begin{equation}
  V_{H^1} = 2 \log \left(\frac{2R}{\delta}\right),
\end{equation}
which is at the same time a universal term.
Substituting this result to the expressions of SRRE and SREE, we can see the first term in Equation~\eqref{eq:SRRE_2DScalar} is exactly the usual \renyi entropy of an interval in 2D complex scalar field:
\begin{equation}
  S_n = \frac{1}{3}\left(1+\frac{1}{n}\right)\log \left(\frac{L}{\delta}\right),
\end{equation}
in which we use \(L = 2R\).
The first term in Equation \eqref{eq:SREE_2DScalar} reproduces the usual entanglement entropy of an interval in 2D CFT:
\begin{equation}
  S = \frac{2}{3}\log \left(\frac{L}{\delta}\right).
\end{equation}
Therefore, as in the 4D case, the leading term in the SRRE \(S_n(q)\) is the unresolved \renyi entropy \(S_n\), and similarly, the leading term in SREE \(S(q)\) is the unresolved entanglement entropy \(S\).

Among subsequent terms, the highest-order divergence is again the logarithmic term \(-\log V\), but with a coefficient \(-1\) different from \(-\frac{1}{2}\) in 4D case. This results from the denominator \(\sqrt{V}\) in \(\tilde{\mathcal{Z}}_n(q)\), which is a result of non-vanishing \(F_{n}'(0)\), as mentioned above.
The universal term in \(S(q)\) arises only from the original entanglement entropy:
\begin{equation}
  S(q)_{\log} = \frac{2}{3} \log \left(\frac{L}{\delta}\right).
\end{equation}
Note also that in Equation~\eqref{eq:SREE_2D}, the coefficient of the logarithmic term is \(-\frac{1}{2}\), which is different from \(-1\) in this case. This is because that result is for \emph{compact} scalar\footnote{Due to its twisted boundary condition, a compact scalar with identification \(\phi \sim \phi+2\pi\) has a different heat kernel than that of a non-compact real scalar. This lies beyond the scope of the present work, in which we discuss only non-compact scalars.}, whereas here we consider a non-compact scalar.
In \cite{DiGiulio:BCFT_to_SRE}, the authors also analyze a non-compact (real) scalar in 2D, focusing on the translation symmetry \(\phi\rightarrow\phi+a\); by contrast, we focus on the U(1) rotational symmetry of the complex scalar in this work.

We can further analyze the \(q\)-dependence of \(S(q)\). The expression \eqref{eq:SREE_2DScalar} shows that \(S(q)\) remains independent of the conserved charge \(q\) up to the constant term. This reproduces the equipartition property. To explicitly show the \(q\)-dependence, we expand the expression up to \(\mathcal{O}(V^{-2})\). This term depends on \(q\) through a \(q^2\) dependence, which is reasonable given that the complex scalar field is symmetric under charge conjugation.
This term also sets a constraint on the validity of the large \(V\) approximation, which gives \(q\ll \log \left(\frac{R}{\delta}\right)\) in 2D scalar case.

In summary, for a conformal free complex scalar field in 2D, the expansion of symmetry-resolved entanglement entropy \(S(q)\) in terms of \(V\) consists of the conventional entanglement entropy \(S\), plus the subleading term \(- \log V\), followed by further subleading terms.
The equipartition property holds up to the \(\mathcal{O}(V)\) term, and breaks down at the next higher-order term.
In Section \ref{sec:structure}, we will show that this is a consequence of the expansion in \(V\), and that the dependence on \(q\) can only appear in terms beyond the constant order.

\paragraph{In sectors with charge \(q \sim V\)}
In our analysis above, we treated the charge \(q\) as a constant while taking the large \(V\) limit, effectively examining small charge sectors in the thermodynamic limit (the limit in which the subsystem size is taken to be large compared to the short distance cutoff \(\delta\)). Alternative charge scaling behaviors can also be considered, which would require asymptotic analysis of different form.
Specifically, for 2D scalar, calculations can be performed in the large \(V\) limit for sectors where \(q \sim V\).

Setting \(q = \sigma V\), where \(\sigma\) represents an \(\mathcal{O}(1)\) quantity, then the transformation from \(\mathcal{Z}_n(\mu)\) to \(\mathcal{Z}_n(q)\) takes the form
\begin{equation}
  \mathcal{Z}_n(q) = \int_{-\pi}^{\pi} \frac{\dd \mu}{2\pi} \mathrm{e}^{V (g_n(\mu) -\ii \mu \sigma)}.
\end{equation}
In the large \(V\) limit, this integral can be evaluated using the method of steepest descents, yielding
\begin{equation}
  Z_n(q)=\frac{2 n \mathrm{e}^{\frac{V}{6 n}}}{V\left(4 \pi ^2 n^2 \sigma ^2+1\right)}.
\end{equation}

Furthermore, we obtain SRRE
\begin{multline}
  S_n(q) = \frac{1}{6}\left(1+\frac{1}{n}\right)V - \log V\\
  + \frac{1}{n-1}\left[-n \log(1+4\pi^2\sigma^2) + \log (1+4n^2 \pi^2 \sigma^2)\right] - \frac{\log n}{n-1} + \log 2 + o(V^0).
\end{multline}
Taking the limit \(n\rightarrow 1\), we obtain SREE
\begin{equation}
  S(q) = \frac{V}{3} - \log V -\frac{2}{4 \pi ^2 \sigma ^2+1}-\log \left(4 \pi ^2 \sigma^2 +1\right) + \log 2 + 1 + o(V^0).
\end{equation}
As evident from these two formulas, for charge sectors with \(q \sim V\), the \(q\)-dependence of SRRE and SREE now shows up in constant order, whereas previously equipartition held at constant order and only broken down beyond this order.
This means equipartition does not hold for charge sectors with sufficiently large \(q\).

\section{Holographic Computation}\label{sec:holo_calc}
In this section, we calculate symmetry-resolved entanglement entropy in higher-dimensional holographic theories. We still employ the conformal mapping method introduced in Section \ref{sec:methods} to transform the density matrix \(\rho_{\mathcal{D}}\) of a spherical region into a thermal density matrix in hyperbolic space, as shown in Equation \eqref{eq:rho_to_thermal}. Through the AdS/CFT correspondence~\cite{Maldacena:AdSCFT,Witten:AdS_and_holography,GKP}, the thermal state in hyperbolic space is further dual to a black hole in the bulk theory~\cite{CHM:towards}. Following the interpretation of \(\mathrm{e}^{\ii \mu Q}\) discussed in Section~\ref{sec:methods}, we need to consider charged black holes in the bulk gravity theory, and \(Z(n\beta_0,\ii\mu/(n\beta_0))\) in Equation \eqref{eq:grandZ} is the grand-canonical partition function with imaginary chemical potential of the dual charged black hole.
Using the classical gravity saddle-point approximation, we approximate the grand-canonical partition function with the on-shell action satisfying the required conditions. After computing the desired charged moments and performing the Fourier transforms, we obtain the symmetry-resolved \renyi entropy and symmetry-resolved entanglement entropy\footnote{We note that charged moments in holographic settings have been investigated using similar techniques in~\cite{Baiguera:shape_deform_charged_Renyi}. They employed the conformal transformation and a dual black hole solution. However, there is a key difference: they examine a quantity called charged \renyi entropy proposed in~\cite{Belin:holo_charged_renyi}, which is a function of \(\mu\), whereas we consider symmetry-resolved entropy, which is a function of charge \(q\) and quantifies entanglement within charge sectors.}.

Consider a CFT that has a gravitational dual description. The dual theory is the Einstein-Maxwell theory in \((d+1)\)-dimensional spacetime, with an action that includes gravitational and electromagnetic terms,
\begin{equation}
  I=\frac{1}{2\lp^{d-1}} \int \dd^{d+1}x\sqrt{-g}\left(\frac{d(d-1)}{L^2}+\mathcal{R}-\frac{\ls^2}{4}F_{\mu\nu}F^{\mu\nu}\right),
\end{equation}
in which the scale \(\ls\) depends on the boundary CFT, and is related to the electromagnetic coupling \(e\) by \(e^2 = 2\lp^{d-1}/\ls^2\). The equation of motion of this action is
\begin{equation}
  \mathcal{R}_{\mu\nu} - \frac{1}{2} \mathcal{R} g_{\mu\nu} - \frac{1}{2}g_{\mu\nu} \frac{d(d-1)}{L^2} - \frac{\ls^2}{4}\left(-\frac{1}{2}F^2 g_{\mu\nu} + 2 F_{\mu\alpha} F_{\nu \beta} g^{\alpha\beta}\right) = 0.
\end{equation}

For dimensions $d\geqslant 3$, the required gravitational solution is a topological black hole with a hyperbolic horizon~\cite{Mann:topo_bh,Birmingham:topo_in_ads}, with the metric given by
\begin{equation}
  \dd s^2 = -f(r) \frac{L^2}{R^2}\, \dd\tau^2 +
  \frac{\dd r^2}{f(r)} + r^2\, \dd\Sigma_{d-1}^2,
\end{equation}
where
\begin{equation}
  f(r) = \frac{r^2}{L^2}-1-\frac{m}{r^{d-2}}+\frac{q^2}{r^{2d-4}},
\end{equation}
and $\dd \Sigma_{d-1}^2 = \dd u^2 + \sinh^2 u \, \dd \Omega_{d-2}^2$ is the metric of hyperbolic space $H^{d-1}$ with unit radius of curvature.

The gauge field solution is given by
\begin{equation}
  A = \left(\sqrt{\frac{2(d-1)}{(d-2)}}\frac{L\,q}{R\ls\, r^{d-2}}-\Phi\right)\dd\tau.
\end{equation}
Here, the chemical potential $\Phi$ is chosen such that the electric potential vanishes at the horizon:
\begin{equation}
  \Phi = \sqrt{\frac{2(d-1)}{(d-2)}}\frac{L\,q}{R\, \ls \rhorizon^{d-2}}.
\end{equation}

The black hole mass\footnote{The relationship between parameter $m$ and black hole mass $M$ is $M = \frac{R}{L}m(d-1)V_{\Sigma}/({2\lp^{d-1}})$.} is related to the horizon radius \(\rhorizon\) through
\begin{equation}
  m = \frac{\rhorizon^{d-2}}{L^2}\left(\rhorizon^2-L^2\right)+\frac{q^2}{\rhorizon^{d-2}}.
\end{equation}
The factor $f(r)$ in the metric is  parameterized by \(m\) and can consequently be written in terms of \(\rhorizon\):
\begin{equation}
  f(r) = \frac{r^2}{L^2} - 1 + \frac{q^2  }{r^{2d-4}} - \left(\frac{\rhorizon}{r}\right)^{d-2} \left(\frac{\rhorizon^2}{L^2}-1 + \frac{q^2}{\rhorizon^{2d-4}}\right).
\end{equation}
The temperature of the black hole is given by
\begin{equation}\label{eq:temperature}
  T = \frac{L}{R}\frac{f'(\rhorizon)}{4\pi}=\frac{T_0}2\left[d\,\frac{\rhorizon}{L}-(d-2)\frac{L}{\rhorizon}\left(1+\frac{d-2}{2(d-1)}\left(\frac{\Phi\,\ls R}{L}\right)^2\right)\right].
\end{equation}

The black hole entropy is proportional to its horizon area:
\begin{equation}\label{eq:bh_entropy}
  S = \frac{2\pi}{\lp^{d-1}} V_{\Sigma} \rhorizon^{d-1}.
\end{equation}
The grand potential $F=-T \log Z$ can be derived from its relationship with entropy:
\begin{equation}\label{eq:entp_to_grand}
  F(T,\Phi) = -\int S(T,\Phi)\dd T.
\end{equation}
Introducing the reduced horizon radius
\begin{equation}
  x = \rhorizon/L,
\end{equation}
the temperature from Equation~\eqref{eq:temperature} can be expressed in terms of \(x\) (and chemical potential $\Phi$) as
\begin{equation}\label{eq:T_xPhi_bh}
  T(x,\Phi) = \frac{T_0}{2}\left[dx-\frac{d-2}{x}k_d(\Phi)\right],
\end{equation}
where, for notational simplicity, we define the function $k_d(\Phi)$ as
\begin{equation}
  k_d(\Phi) = 1+\frac{d-2}{2(d-1)}\left(\frac{\Phi\,\ls R}{L}\right)^2.
\end{equation}
The reduced horizon radius $x$ can then be expressed in terms of temperature \(T\) and \(\Phi\):
\begin{equation}
  x(T,\Phi) = \frac{T}{d\, T_0}+\sqrt{\left(\frac{T}{d\, T_0}\right)^2+\frac{d-2}{d}k_d(\Phi)}.
\end{equation}
The integration of the grand potential in Equation~\eqref{eq:entp_to_grand} yields\footnote{When choosing $x$ and $\Phi$ as independent variables, the entropy does not depend on $\Phi$, as evident from the entropy formula \eqref{eq:bh_entropy}.}
\begin{equation}
  \begin{aligned}
    F(x,\Phi) & = - \int S(x,\Phi) \frac{\pd T (x,\Phi)}{\pd x} \dd x                                    \\
              & = - \pi V_{\Sigma}T_0 \left(\frac{L}{\lp}\right)^{d-1} \left(x^d+x^{d-2} k(\Phi)\right).
  \end{aligned}
\end{equation}

To perform the computation of SREE, the thermodynamic variables such as \(x\) and \(\Phi\) need to be expressed in terms of $n$ and \(\mu\) in the context of entanglement entropy via Equation~\eqref{eq:T_n_relation} and \eqref{eq:T_xPhi_bh}, and the relationship between \(\Phi\) and \(\mu\)
after analytical continuation,
\begin{equation}
  \Phi = \mu T.
\end{equation}
The grand-canonical partition function with imaginary chemical potential can be expressed as
\begin{equation}
  Z(n,\ii\mu) = \tr\left[\mathrm{e}^{-nH/T_0+\ii \mu Q}\right] = \mathrm{e}^{V g_n(\mu)},
\end{equation}
in which the expression of \(g_n(\mu)\) is as follows,
\begin{multline}\label{eq:gnmu}
  g_n(\mu) = \pi  n  \left( \frac{L}{\lp} \right)^{d-1} \left(\frac{n \sqrt{(d-2) d \left(1-\frac{(d-2) \mu ^2 \ls^2}{8 \pi ^2 (d-1) L^2 n^2}\right)+\frac{1}{n^2}}+1}{d n}\right)^d  \\
  \left(\frac{d^2 n^2 \left(1-\frac{(d-2) \mu ^2 \ls^2}{8 \pi ^2 (d-1) L^2 n^2}\right)}{\left(n \sqrt{(d-2) d \left(1-\frac{(d-2) \mu ^2 \ls^2}{8 \pi ^2 (d-1) L^2 n^2}\right)+\frac{1}{n^2}}+1\right)^2}+1\right).
\end{multline}
Despite the long expression, \(g_n(\mu)\) can take simple values at special points, e.g.,
\begin{equation}
  g_1(0) = 2\pi \left( \frac{L}{\lp}\right)^{d-1}.
\end{equation}

In the large \(V\) limit, the inverse Fourier transform in Equation \eqref{eq:tildeZnq} can be asymptotically expanded to yield
\begin{equation}
  \tilde{\mathcal{Z}}_n(q) = \frac{e^{V g_n(0)}}{\sqrt{2 \pi } \sqrt{-V g^{''}_{n}(0)}} \left(1 + \frac{4 q^2 g^{''}_{n}(0)+g^{(4)}_{n}(0)}{V g^{''}_{n}(0)^2}\right),
\end{equation}
where the function \(g_n(\mu)\) is defined in Equation \eqref{eq:gnmu}. From this, we can obtain the symmetry-resolved \renyi entropy, and by taking the limit \(n \rightarrow 1\), we derive the symmetry-resolved entanglement entropy
\begin{equation}\label{eq:SREE_holo}
  S(q) = 2\pi \left( \frac{L}{\lp}\right)^{d-1} V - \frac{1}{2} \log V + h_0 + h_{-1},
\end{equation}
where \(h_0\) and \(h_{-1}\) represent terms of order \(V^0\) and \(V^{-1}\) in the expansion, respectively:
\begin{equation}
  \begin{aligned}
    h_0    & = \frac{1}{4} \left(\frac{3-2 d}{d-1}-\log \left(\frac{d-2}{2} \left(\frac{\ls}{L}\right)^2 \left(\frac{L}{\lp}\right)^{d-1}\right)\right)                     \\
    h_{-1} & = \frac{1}{V}\, \left(\frac{L}{\ls}\right)^2 \left(\frac{\lp}{L}\right)^{d-1} \frac{\left(32 \pi ^2 (d-1)^3 q^2+3 d (d-2)^3 \ls^2/L^2\right)}{2 \pi  (d-1)^4}.
  \end{aligned}
\end{equation}

We can analyze the terms in \(S(q)\) as follows. First, the leading term in the symmetry-resolved entanglement entropy \(S(q)\) is the unresolved entropy \(S\), which corresponds to the entropy of an uncharged hyperbolic black hole in holographic gravity\footnote{We should note that the symmetry-resolved entanglement entropy \(S(q)\) is \emph{not} equal to the entropy of a charged hyperbolic black hole.}. This term represents the entanglement entropy of the chosen spherical region in the dual field theory~\cite{CHM:towards,Hung:holo_renyi,Belin:holo_charged_renyi}. Therefore, the leading term in the symmetry-resolved entanglement entropy is proportional to the hyperbolic space volume \(V\), with its leading contribution manifesting the area law. This characteristic aligns with the field theory computations presented in Section \ref{sec:free_calc}.

The next-order term is the logarithmic contribution \(-\frac{1}{2} \log V\), which introduces new universal terms beyond those present in the original entanglement entropy. For odd \(d\), the universal term in \(S(q)\) takes the form
\begin{equation}
  S(q)_{\mathrm{univ}} = - \frac{d-2}{2} \log \left(\frac{R}{\delta}\right).
\end{equation}
The original constant universal term \(\frac{\pi^{d/2}}{\Gamma(d/2)}(-1)^{\frac{d-1}{2}}\) in the hyperbolic space volume \(V\) loses its universal character, as transformations like \(\delta \rightarrow a\delta\) generate new contributions from the logarithmic term, modifying the constant term. For even \(d\), both leading terms contribute to universal terms:
\begin{equation}
  S(q)_{\mathrm{univ}} = 2\pi \left( \frac{L}{\lp}\right)^{d-1} V_{\mathrm{univ}} - \frac{d-2}{2} \log \left(\frac{R}{\delta}\right).
\end{equation}
In the limit with large degrees of freedom \(\left( \frac{L}{\lp}\right)^{d-1} \rightarrow \infty\), the first term dominates the second, leading to an approximate universal term in the symmetry-resolved entanglement entropy \(S(q)\):
\begin{equation}
  S(q)_{\mathrm{univ}} \simeq 2\pi \left( \frac{L}{\lp}\right)^{d-1} \frac{\pi^{d/2}}{\Gamma(d/2)} \times
  \begin{cases}
    (-)^{\frac{d}{2}-1} \frac{2}{\pi} \log (2 R / \delta), & d \, \text{even} \\
    (-)^{\frac{d-1}{2}},                                   & d \, \text{odd}
  \end{cases}
\end{equation}
where we have used the expression \eqref{eq:V_univ} for the universal term in the hyperbolic space volume.

We can also analyze the dependence of \(S(q)\) on \(q\). The expression for SREE \eqref{eq:SREE_holo} shows that \(S(q)\) remains independent of the conserved charge \(q\) up to the constant term, thus demonstrating the equipartition property in holographic theories for general dimensions. The dependence on \(q\) first appears in \(h_{-1}\) and takes the form of \(q^2\). This quadratic behavior reflects the symmetry of the charged theory under charge conjugation.

In summary, for holographic field theories, the SREE \(S(q)\) exhibits an expansion structure in \(V\) consisting of the unresolved entanglement entropy \(S\), followed by a subleading term of order \(\log V\), and a constant term. All three terms are independent of \(q\), demonstrating the equipartition property up to the constant order. The \(q\)-dependence shows up at higher-order terms.
This structure aligns with both the two-dimensional free compact scalar results in Equation \eqref{eq:SREE_2D} and the higher-dimensional free field theory computations presented in Section \ref{sec:free_calc}.
In Section \ref{sec:structure} we will show that this structure and equipartition is a consequence of the expansion in \(V\), and any violation of the equipartition property can only appear in terms beyond the constant order.

\section{Universal Structure of SREE}\label{sec:structure}

In this section, we explore the expansion structure of symmetry-resolved entanglement entropy. In Section \ref{sec:structure_summary}, we synthesize results of SREE in two dimensions and higher dimensions presented in Section \ref{sec:free_calc} and Section \ref{sec:holo_calc} to summarize its universal expansion structure.
In Section \ref{sec:structure_proof}, we conduct a systematic asymptotic analysis for systems with \(\mathrm{U}(1)\) global symmetry, deriving universal expressions for the charged moments \(\mathcal{Z}_n(\mu)\), symmetry-resolved partition functions \(\mathcal{Z}_n(q)\), symmetry-resolved \renyi entropy \(S_n(q)\), and symmetry-resolved entanglement entropy \(S(q)\). These expressions reveal the universal structure of SREE and directly demonstrate its equipartition property up to constant order.

\subsection{Expansion structure of SREE}\label{sec:structure_summary}
First, let us review the SREE in two-dimensional field theories in previous studies and higher-dimensional results discussed in previous sections. Taking the massless compact scalar field as an example~\cite{Xavier:Equipartition,Goldstein:SRE}, the SREE for a line segment of length \(L\) is given by
\begin{equation}\label{eq:SRRE_2D_copy}
  S(q) = S - \frac{1}{2} \log \left(\log L\right) - \frac{1}{2} \log\left(\frac{2K}{\pi}\right) -\frac{1}{2} + o(L^0),
\end{equation}
where \(K\) parametrizes the compactification radius of the compact scalar.
Analogously, the SREE results in Section \ref{sec:free_calc} and Section \ref{sec:holo_calc} reveal the following structure:
\begin{equation}
  S(q) = S - \# \log V + h_{0} + h_{-1} + h_{-2} + \cdots,
\end{equation}
where \(h_{n}\) represents order \(V^{n}\) terms in the expansion.
The structure of SREE exhibits several common key features: (1) the leading term is the entanglement entropy; (2) the subleading correction is of order \(\log V\) (recall that \(V \sim \log L\) in 2D); (3) the equipartition property holds up to constant order \(h_{0}\).
These structural properties have been discovered through computations in spin chains~\cite{Xavier:Equipartition}, conformal field theories~\cite{Goldstein:SRE}, and holographic methods~\cite{Zhao:SRE_in_AdSCFT}.

\subsection{Expansion structure via asymptotic analysis}\label{sec:structure_proof}
Recall that the SREE \(S(q)\) is derived from \(\mathcal{Z}_n(q)\), which is obtained through an inverse Fourier transform of \(\mathcal{Z}_n(\mu)\). In asymptotic analysis for large hyperbolic space volume \(V\), both \(\mathcal{Z}_n(q)\) and \(S(q)\) are expressed as expansions in powers of \(V\), allowing for a systematic examination of their structures.
This process has been done in the last two sections for specific theories.
In this section, we will perform the analysis from a purely mathematical perspective, demonstrating the structure and the equipartition property of SREE.

We restate the transform here:
\begin{equation}
  \mathcal{Z}_n(q) = \int_{-\pi}^{\pi} \frac{\dd \mu}{2\pi } \mathrm{e}^{-\ii \mu q} \mathcal{Z}_n(\mu) = \int_{-\pi}^{\pi} \frac{\dd \mu}{2\pi } \mathrm{e}^{-\ii \mu q} \mathrm{e}^{V g_n(\mu)}.
\end{equation}
In the asymptotic analysis for this integral transform, both functions \(\mathrm{e}^{-\ii\mu q}\) and \(\mathrm{e}^{V g_n(\mu)}\) need to be expanded around the maximum point of \(g_n(\mu)\) at \(\mu=c\) (specifically, \(c=0\) in the calculations of the previous section)~\cite{Bender:asymp}. The leading behavior of the function \(g_n(\mu)\) near its maximum is controlled by its first few derivatives. If \(g'_n(\mu)=0\) at the maximum point \(\mu=0\) (for 4D scalar in Section \ref{sec:free_calc_4DScalar} and holographic theories in Section \ref{sec:holo_calc}), then beyond the function value \(g_n(0)\) itself, its leading behavior is controlled by the second derivative; if \(g'_n(\mu)\ne 0\) (as for 2D scalar in Section \ref{sec:free_calc_2DScalar}, where \(g'_n(\mu)\) is in fact discontinuous at \(\mu=0\)), then after the function value, the next most important term is the first derivative. Below we examine both cases.

\paragraph{Case 1: \(F'(\mu)=0\) at \(\mu=0\)}
In this case, the general expression for the Fourier integral result, accurate up to the second order, is given by\footnote{In the computation for the four-dimensional free scalar field, \(g_n(\mu)\) reaches its maximum at \(\mu=0\), but it is not fourth-order differentiable at this point. Thus, the asymptotic expansion needs to be applied separately for \(\mu<0\) and \(\mu>0\). However, the forms of the expansion results are the same, so no special distinction is made.}
\begin{equation}\label{eq:SRpart_2nd_order}
  \mathcal{Z}_n(q) = \frac{ \ee^{V g} }{\sqrt{2\pi}} \frac{1}{\sqrt{V (-g'')}} \left(h + \frac{-\frac{h''}{2 g''}+\frac{g^{(3)} h'}{2 (g'')^2}-\frac{5 h \left(g^{(3)}\right)^2}{24 (g'')^3}+\frac{h g^{(4)}}{8 (g'')^2}}{V}\right).
\end{equation}
Here, \(h(\mu) = \mathrm{e}^{-\ii\mu q}\), \(g^{(k)}\) denotes the \(k\)-th derivative with respect to \(\mu\), and both the function \(h\) and the derivatives of \(g\) are evaluated at \(\mu = c\) (\(c=0\) for the cases we consider in this paper). The reason that it's \(\sqrt{V}\) in the denominator of \(\mathcal{Z}_n(q)\) is because the second-derivative term is the primary term after the function value. 
The derivation and explanation of this result are discussed in Appendix \ref{sec:higher_asymp}.

Let's write the above expression formally as
\begin{equation}\label{eq:Znq_formal}
  \mathcal{Z}_n(q) = \frac{\mathrm{e}^{Vg_n(c)}}{\sqrt{2\pi} \sqrt{V f_n}} \left(1+\frac{a_n(q)}{V}\right),
\end{equation}
where \(f_n=-g_{n}''(0)>0\), and \(h=1\) for the cases we consider in this paper.
Substituting it into the expression for SRRE, we obtain
\begin{multline}\label{eq:SRRE_general}
  S_n(q) = V\frac{(n g_1(c)-g_n(c))}{n-1} - \frac{1}{2} \log V \\
  + \frac{-n \log f_1+\log f_n}{2 (n-1)} - \frac{1}{2}\log (2\pi) + \frac{1}{V}\frac{n a_1(q)-a_n(q)}{n-1}.
\end{multline}
Taking the limit \(n\rightarrow 1\), we then obtain SREE \(S(q)\):
\begin{multline}\label{eq:SREE_general}
  S(q) = V \left(g_1(c)-\left.\frac{\pd g_n(\mu)}{\pd n}\right|_{n=1,\mu=c}\right) - \frac{1}{2} \log V +\frac{f'(1)}{2 f(1)}-\frac{1}{2} \log (f(1)) -\frac{1}{2}\log (2\pi) \\
  + \frac{1}{V} \left(a_1(q)- \left.\frac{\pd a_n(q)}{\pd n}\right|_{n=1}\right).
\end{multline}
These two equations conform to the expansion structure described in Section \ref{sec:structure_summary}: the first term is the unresolved \renyi entropy and entanglement entropy; the second term is of order \(\log V\); the next term is constant, and SREE remains independent of \(q\) up to this order, exhibiting the equipartition property; the breaking of equipartition appears at order \(V^{-1}\).

\paragraph{Case 2: \(F'(\mu)\ne0\) at \(\mu=0\)}
In this case, the first-derivative term is the most important term after the function value at \(\mu=0\).
The general expression for the Fourier integral result is approximately given by\footnote{The derivative is actually discontinuous at \(\mu=0\). However, as explained in the previous footnote, to perform this asymptotic expansion we treat the cases \(\mu=0^+\) and \(\mu=0^-\) separately and then sum the results. the discontinuity presents no new conceptual difficulties.}
\begin{equation}\label{eq:SRpart_3rdOrder}
  \mathcal{Z}_n(q) = \frac{\mathrm{e}^{V g}}{\pi V(-g')} \left(1+ \frac{g'' }{V g'^2} + \frac{3 g''^2 -q^2 g'^2}{V^2 g'^4} \right),
\end{equation}
with all functions being evaluated at \(\mu = c=0\). Note that it's now \(V\) in the denominator rather than \(\sqrt{V}\), and the reason is the leading derivative term is a first-derivative.
Note that we keep terms in three orders to include the \(q\)-dependence.

Let's write the above expression formally as
\begin{equation}\label{eq:Znq_formal_ver2}
  \mathcal{Z}_n(q) = \frac{\mathrm{e}^{V g_n}}{\pi V f_n}\left(1 + \frac{a_n}{V} + \frac{b_n(q)}{V^2}\right)
\end{equation}
where \(f=-g'(0)>0\).
From this we can obtain SRRE:
\begin{multline}
  S_n(q) = V\frac{(n g_1(c)-g_n(c))}{n-1} - \log V  + \frac{-n \log f_1+\log f_n}{n-1} - \log(\pi)\\
  + \frac{1}{V} \frac{n a_1-a_n}{(n-1)} + \frac{1}{V^2} \frac{-n a_1^2+a_n^2+2 n b_1(q)-2 b_n(q)}{2 (n-1)}
\end{multline}
Taking the limit \(n\rightarrow 1\),  we then obtain SREE:
\begin{multline}\label{eq:SREE_general_ver2}
  S(q) = V \left(g_1(c)-\left.\frac{\pd g_n(\mu)}{\pd n}\right|_{n=1,\mu=c}\right) - \log V +\frac{f'(1)}{f(1)} - \log (\pi f(1)) \\
  + \frac{1}{V} \left(a_1- \left.\pd_n a_n\right|_{n=1}\right) - \frac{a_1^2 - 2b_1(q)-2a_1 \left.\partial_n a_n\right|_{n=1}+2\left.\partial_n b_n(q)\right|_{n=1}}{2V^2}.
\end{multline}
This conforms to the expansion structure described in Section \ref{sec:structure_summary}---the first term is the unresolved entanglement entropy; the second term is \(-\log V\); the next term is constant, and the SREE remains independent of \(q\) up to this order, exhibiting the equipartition property; the breaking of equipartition appears at order \(V^{-1}\).
Note that \(a_n=2n\) in 2D scalar in Section \ref{sec:free_calc_2DScalar}, thus the coefficient of order \((V^{-1})\), i.e. \(a_1- \left.\pd_n a_n\right|_{n=1}\), vanishes in that case.

Therefore, we have demonstrated that for theories with \(\mathrm{U}(1)\) global symmetry, when expanded in terms of \(V\), the symmetry-resolved entanglement entropy exhibits the structure described in Section \ref{sec:structure_summary}. The equipartition property holds up to order \(V^0\), with violations first appearing at higher orders. The asymptotic analysis of charged moments in this section reveals that both the expansion structure of symmetry-resolved entanglement entropy and the equipartition property emerge from the expansion in \(V\), independent of specific details of the theory.

\section{Discussion}\label{sec:discussion}
This paper presents a method for computing symmetry-resolved entanglement entropy of spherical regions in higher-dimensional CFT. We demonstrate the method through explicit calculations in free field theories and holographic field theories. We summarize the universal expansion structure of symmetry-resolved entanglement entropy, demonstrating that this structure necessarily emerges in the large subsystem size limit, while also demonstrating the equipartition property up to constant order.

This conformal mapping technique is applicable across general dimensions. This technique has been used in entanglement entropy computations, such as verifying the holographic entanglement entropy formula~\cite{CHM:towards}, understanding \renyi entropy properties in free field theories~\cite{Klebanov:renyi_free}, and studying \renyi entropy in holographic duality~\cite{Hung:holo_renyi}, consistently working across various dimensions. To investigate SREE in higher dimensions, we appropriately extended this computational technique to calculate charged moments, enabling the computation of SREE for spherical regions in arbitrary dimensions. According to this method, computing entanglement entropy or SREE reduces to solving thermodynamics on the corresponding hyperbolic space.
In Section \ref{sec:free_calc}, since the thermodynamics of free field theory on hyperbolic space can be solved exactly, we present a complete computation of SREE. This method clearly extends to general dimensions; although heat kernel computations on hyperbolic space become more complex with increasing dimensions~\cite{Grigor'yan:heatKer_hyperbolic}, it only presents computational rather than theoretical challenges.
Furthermore, in Section~\ref{sec:holo_calc}, we discuss holographic field theory using a bottom-up approach, employing the classical gravity approximation to analyze the thermodynamics of dual black holes, thereby computing charged moments and SREE. Since black hole thermodynamic quantities have expressions applicable across dimensions, the holographic computations also reveal SREE results applicable in general dimensions.

By comparing the results from free field theory and holographic field theory, we find that the SREE (and SRRE in intermediate steps) exhibits the same structure in both theories: the leading term is the unresolved entanglement entropy, the second term is logarithmic in the first term, and terms up to constant order are independent of the charge, displaying the equipartition property.
These structures and properties also align with those observed in models in previous studies.
Based on these observations, in Section~\ref{sec:structure}, we provide a proof for systems with \(\mathrm{U}(1)\) global symmetry, showing that the structure of SREE and the equipartition property are general results of expansion in subsystem size (see Equations~\eqref{eq:SREE_general} and \eqref{eq:SREE_general_ver2}), independent of specific theoretical details.
Our proof reveals the conditions for the existence of universal structure in SREE and demonstrates that the equipartition property is essentially an approximate behavior in the large subsystem size limit.

We briefly outline potential directions for future research based on this work.

A natural extension is the study of SREE with non-Abelian global symmetries. In two-dimensional field theory, discussions of SREE in non-Abelian WZW models have already emerged~\cite{Calabrese:SREE_in_WZW}, revealing that, while equipartition still holds at the leading order in the expansion of subsystem length \(L\), terms related to representation dimensions appear at the constant order, breaking equipartition, which is a novel feature of the non-Abelian case.
However, the study of non-Abelian SREE in higher dimensions remains unexplored. We expect to extend the conformal mapping method presented in this work to obtain results of non-Abelian SREE in higher dimensions.
Comparing these results with the \(\mathrm{U}(1)\) case would allow us to analyze new properties that might arise from non-commutativity.

The proof in Section \ref{sec:structure} demonstrates that the equipartition property in the leading orders of SREE emerges as an approximation when choosing \(V\) as the expansion parameter, independent of specific theory details.
As explained at the end of Section~\ref{sec:methods}, using \(V\) as a large expansion parameter is equivalent to expansion in a large subsystem size, which is a standard regularization method in studying quantum field theory entanglement entropy. However, introducing the conserved charge \(q\) as a new parameter imposes additional requirements on this expansion. For instance, the perturbative validity of the \(V\)-expansion of SREE requires conditions such as \( q^2 \ll V \), as evident from earlier derivations.
Future investigations should focus on exploring the applicability and limitations of the \(V\)-expansion method.
As an initial exploration, in Section~\ref{sec:free_calc_2DScalar}, we analyze a regime where \( q \sim V \), revealing qualitatively distinct features in the SREE compared to the \( q^2 \ll V \) case.
Additionally, it would be valuable to investigate what new physical properties might emerge when choosing different expansion parameters in other limits (such as large \(q\) expansion when \(V\) is considered as a fixed finite quantity). The expansion structure and universal terms in such expansions might reveal properties that are not apparent in the \(V\)-expansion.

\acknowledgments

We thank Yuhan Fu, Shunji Matsuura, Ren{\'e} Meyer, Zhi Wang, and Suting Zhao for helpful discussion. This work is supported by NSFC grant 12375063 and is also sponsored by Shanghai Talent Development Fund. YZ would like to thank the organizers of "Non-perturbative methods in QFT" in Kyushu University for hosting.

% \paragraph{Note added.} This is also a good position for notes added
% after the paper has been written.

\appendix

\section{Review of Symmetry-Resolved Entanglement}\label{sec:SRE}
In this appendix, we present a motivation for symmetry-resolved entanglement. Beginning with entropy coarse-graining from Shannon's original thoughts, we extend these principles to the quantum domain to develop the concept of symmetry-resolved entanglement.

\subsection{Entropy and the coarse-graining property}

Shannon entropy is used to measure the uncertainty in a set of possible events \(1,\cdots, n\) with some probability distribution \(\{p_1,\cdots, p_n\}\), and is given by~\cite{Shannon}
\begin{equation}
  S(\{p_i\}) = -\sum_{i} p_i \log p_i.
\end{equation}
In the language of thermodynamics, when a macrostate contains many microstates with their own probabilities, an entropy can assign to the macrostate according to Shannon's formula.

\begin{figure}[tbh]
  \centering
  \includegraphics{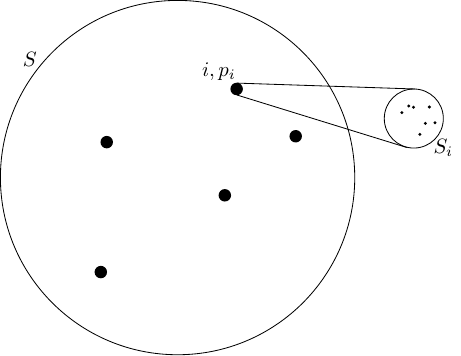}
  % We can also generate this image by TikZ
  % \begin{tikzpicture}
  %   \draw (0,0) circle (3cm);
  %   \node at (-2.5,2) {$S$};
  %   \fill (1,1.5) circle (3pt);
  %   \node at (0.7, 1.8) {\(i,p_i\)};
  %   \fill (2,0.7) circle (3pt);
  %   \fill (0.8,-0.3) circle (3pt);
  %   \fill (-1.3,-1.6) circle (3pt);
  %   \fill (-1.2,0.6) circle (3pt);

  %   \draw (4,1) circle (0.5cm);
  %   \node at (4.5,0.5) {$S_i$};
  %   \foreach \i in {1,...,7}{
  %       \pgfmathsetmacro{\angle}{rand*360}
  %       \pgfmathsetmacro{\x}{4 + 
  %       rand*0.5*cos(\angle)}
  %       \pgfmathsetmacro{\y}{1 + rand*0.45*sin(\angle)}
  %       \fill (\x,\y) circle (0.8pt);
  %     }

  %   \draw (1,1.6) -- (4.02,1.5);
  %   \draw (1,1.4) -- (3.88,0.51);

  % \end{tikzpicture}
  \caption{Illustration for microstates (big black dots) and sub-microstates (small dots in small circle). The \enquote{microstate} \(i\) appears with probability \(p_i\) and has its own entropy \(S_i\). The actual total entropy comes from uncertainty of probability distribution \(\{p_i\}\) and uncertainty within state \(i\).}
  \label{fig:coarse-graining}
\end{figure}

However, if the microstates can be further broken down into what we dub \enquote{submicrostates}, there will be more uncertainty.\footnote{That is to say, those microstates are actually not the most detailed description of the system. This is in fact natural in physics if we take the viewpoint of effective theory.} See Figure \ref{fig:coarse-graining} for an illustration.
This additional uncertainty is the average of entropies of the submicrostates. Then the actual total entropy is
\begin{equation}
  \label{eq:coarse-graining}
  S = -\sum_{i}p_i\log p_i + \sum_i p_i S_i.
\end{equation}
Given Shannon's definition of entropy, this formula can be proved. However, it's worth noting that Shannon, taking the opposite route, took (\ref{eq:coarse-graining}) as one of the defining properties to \emph{derive} his expression of entropy \cite{Shannon}. Note that the argument of total entropy $S$ now is more detailed probability distribution, which we don't explicitly write down. Also note that, at a deeper level, there might be even more uncertainty in \(S_i\), and \(S_i\) can be decomposed in the same way as \(S\) in (\ref{eq:coarse-graining}). In this sense, \(S_i\) and \(S\) are in fact of the same nature. This will still be true in quantum physics, where both \(S_i\) and \(S\) are entanglement entropy as discussed below in the next subsection, while \(-\sum_{i}p_i\log p_i \) is still classical Shannon entropy.

\subsection{In quantum theory: symmetry-resolved entanglement}
In quantum theory, the natural generalization of Shannon entropy is von Neumann entropy:
\begin{equation}
  S(\rho) = -\tr[\rho\log\rho],
\end{equation}
where \(\rho\) is a density matrix. The analog of the decomposition to sub-microstates in the previous subsection is the decomposition of the density matrix into a direct sum, \(\rho = \bigoplus_q \tilde{\rho}(q)\), which can be realized in entanglement with global symmetry.

Consider a bipartite system with global U(1) symmetry with conserved charge \(Q\), which is a sum of charge of the two subsystems, \(Q=Q_A+Q_B\). When the total system density matrix \(\rho\) commutes with the conserved charge \(Q\), it can be proved that, by partial trace, the subsystem density matrices still commute with the corresponding charge operator, i.e., \([\rho_A,Q_A] = [\rho_B,Q_B]=0\). Therefore, the subsystem density matrix  \(\rho_A\) (let's focus on subsystem \(A\) from now on) will be block-diagonal in the eigenbases of the conserved charge \(Q_A\):
\begin{equation}
  \rho_A = \bigoplus_q \tilde{\rho}_A(q).
\end{equation}
The probability of finding eigenvalue \(q\) in a measurement of \(Q_A\) is given by
\begin{equation}
  p_A(q) = \tr[\tilde{\rho}_A (q)].
\end{equation}
Then \(\rho_A\) can also be written as
\begin{equation}
  \rho_A = \bigoplus_q p_A(q) \rho_A(q),
\end{equation}
where \(\rho_A(q)\) is now normalized, i.e. \(\tr[\rho_A(q)] = 1\).

Following the idea of coarse-graining of entropy in the last subsection, a quantum-mechanical counterpart of \(S_i\) will naturally be given by the von Neumann entropy of the (normalized) density matrices \(\rho_A(q)\) in charge sectors, called symmetry-resolved entanglement entropy (SREE), defined as
\begin{equation}
  S(q) = -\tr[\rho_A(q) \log \rho_A(q)].
\end{equation}
Similarly, the symmetry-resolved \renyi entropy is defined as
\begin{equation}
  S_n(q) = \frac{1}{1-n} \log \tr[\rho_A(q)].
\end{equation}

Analogous to the coarse-graining property \eqref{eq:coarse-graining} of Shannon entropy, the total von Neumann entropy of \(\rho_A\) can also be decomposed into two parts as follows:
\begin{equation}
  S(\rho_A) = -\sum_{q}p_A(q) \log p_A(q) + \sum_q p_A(q) S(\rho_A(q)).
\end{equation}

\section{Heat Kernel Essentials}\label{sec:heat_kernel_essentials}
In this section, we present a concise, pedagogical introduction to the essentials of heat kernel in QFT needed for the main text, and derive key expressions, such as the heat kernel with twisted boundary condition on \(S^1\) (Equation \eqref{eq:K_S1_mu}).

\subsection{Basic definition of heat kernel}
The heat kernel $K(x, y, t)$ is originally defined as the fundamental solution to the heat equation:
\begin{equation}
  \left(\partial_t + D_x\right) K(x, y, t) = 0,
\end{equation}
with the initial condition as a point source
\begin{equation}
  K(x, y, 0) = \delta(x - y),
\end{equation}
where $D_x$ is a differential operator (often the Laplacian or Laplace-type operator) acting on the $x$ variable.

In QFT, the heat kernel is a powerful tool for evaluating functional determinants and partition functions. As an example of heat kernel in QFT, we consider the (Euclidean) path integral representation of the generating functional for real scalar field $\phi$:
\begin{equation}
  Z = \int \mathcal{D}\phi \, \exp(-S[\phi])
\end{equation}
where the action $S$ only contains the quadratic part (free field):
\begin{equation}
  S = \frac{1}{2}\int\phi D \phi \sqrt{g}\dd^d x 
\end{equation}
Here, \(D\) is the differential operator
\begin{equation}
  D = D_0 := -\nabla^\mu \nabla_\mu.
\end{equation}

The heat kernel can be expressed as:
\begin{equation}\label{eq:K_transition}
  K_D(x, y, t) := \langle x | \ee^{-tD} | y \rangle.
\end{equation}
One can easily verify that \(K(x, y, t)\) satisfies the heat equation with the initial condition being a \(\delta\) function source:
\begin{equation}
  \begin{cases}
    (\partial_t + D_x) K_D(x, y, t) = 0, \\
    K_D(x, y, 0) = \delta(x, y).
  \end{cases}
\end{equation}
The heat kernel solution to this equation depends on the differential operator \(D\) and the boundary condition.

For the complex scalar we consider in Section~\ref{sec:free_calc}, the path integral can be evaluated by a Gaussian integral, yielding
\begin{equation}
  Z \sim \frac{1}{\det D}.
\end{equation}
The free energy \(F = -\log Z\) can be computed by~\cite{Vassilevich:heatkernel_manual}
\begin{equation}
  F = -\log Z = \log \det D =  - \int_0^{\infty} \frac{\dd t}{t} \, \tr\left[\ee^{-tD}\right].
\end{equation}
Using expression \eqref{eq:K_transition}, we can finally write \(F\) in terms of the heat kernel:
\begin{equation}
  F = -\int_{0}^{\infty}\frac{\dd t}{t}\int \dd^n x\sqrt{g} K_D(x,x,t).
\end{equation}
Therefore, the computation of the generating functional (and charged moments needed in this paper) for the free field can be completed using heat kernel techniques. The specific form of the heat kernel depends on the differential operator \(D\) and any imposed boundary condition (in this paper, twisted boundary conditions on the Euclidean time circle).

We also note the connection to the real (non-compact) scalar. A free complex scalar field can be viewed as two independent real scalars, so its action decomposes into the sum of two identical real-scalar actions. Evaluating again the Gaussian integral yields partition function \(Z_{\text{real}}\sim \frac{1}{\sqrt{\det D}}\). Hence, for \(\mu=0\), the free energy of complex scalar is exactly twice that of that of real scalar: \(F_{\text{complex}} = 2F_{\text{real}}\).

The above derivation of the heat kernel is rather schematic; for a more rigorous mathematical treatment and its applications in QFT, see, for example, \cite{Vassilevich:heatkernel_manual}.

\subsection{Heat Kernel on \(S^1_\beta\) with twisted boundary condition}\label{sec:heat_kernel_essentials_derivation}
Here we derive the heat kernel of the Laplace operator on \(S^1\) in Equation \eqref{eq:K_S1_mu} with twisted boundary condition \eqref{eq:twisted_BC}.

Starting from the heat kernel as a \enquote{transition amplitude} \eqref{eq:K_transition}, by inserting an orthonormal basis in it, we obtain the heat kernel as a \enquote{mode sum}:
\begin{equation}\label{eq:mode_sum}
  K(x,y,t) = \sum_n \Braket{x|\phi_n} \Braket{\phi_n|\ee^{-tD}|y} = \sum_n \phi_n(x) \phi_n^*(y) \ee^{-t\lambda_n},
\end{equation}
where \(\phi_n\) are eigenfunctions of \(D\) with eigenvalues \(\lambda_n\).

Consider a circle \(S^1_\beta\) of circumference \(\beta\), and a field \(\phi(x)\) satisfying the twisted boundary condition:
\begin{equation}
  \phi(x + \beta) = \ee^{\ii \mu} \phi(x).
\end{equation}
The corresponding eigenfunctions and eigenvalues are
\begin{equation}
  \phi_n(x) = \frac{1}{\sqrt{\beta}}\ee^{\ii \frac{2\pi n+\mu}{\beta}x},\, \lambda_n = \left(\frac{2\pi n+\mu}{\beta}\right)^2, \, n \in \mathbb{Z}.
\end{equation}
Substituting this into the mode sum and using the Poisson summation formula to convert it to an \enquote{image sum}, we obtain
\begin{equation}
  K_{S^1_\beta}(x_1, x_2, t) = \frac{1}{\sqrt{4\pi t}} \sum_{k \in \mathbb{Z}} \exp\left(-\frac{(x_1 - x_2 - \beta k)^2}{4t} + \ii k \mu \right).
\end{equation}
By setting \(\beta = 2\pi n\), this expression recovers Equation \eqref{eq:K_S1_mu}.

\section{Asymptotic Analysis of Fourier Integral}\label{sec:higher_asymp}

This appendix provides the explanation and derivation of the asymptotic expansion formula~\eqref{eq:SRpart_2nd_order} used in Section \ref{sec:structure} and intermediate computational steps in Sections~\ref{sec:free_calc} and~\ref{sec:holo_calc}, for cases in which \(F'(\mu)=Vg'(\mu)=0\) at the maximum point \(\mu=c\). For cases in which \(g'(\mu)\ne 0\), the derivation is similar, and the result is shown in Equation~\eqref{eq:SRpart_3rdOrder}.

In the asymptotic analysis for integrals of the following form, the Laplace method is often used:
\begin{equation}
  I(V) = \int_{a}^{b} h(\mu) \mathrm{e}^{V g(\mu)} \dd \mu.
\end{equation}
Here, it is assumed that \(h(\mu)\) and \(g(\mu)\) are continuous real functions\footnote{In Section \ref{sec:structure}, \(h(\mu)\) is actually a complex function \(h(\mu) = \mathrm{e}^{-\ii\mu q}\), but the integral can be split into two real integrals, which does not affect the applicability of the method discussed in this section.}.

The basic idea of the Laplace method is that the dominant contribution to the integral comes from the neighborhood around the maximum of the (continuous) integrand function~\cite{Bender:asymp}. If, as the parameter \(V\rightarrow \infty\), the integrand forms a sharp peak near its maximum at \(\mu=c\), then a small neighborhood around this point can provide a highly accurate estimate of the integral.
The estimation error is controlled by the limit as \(V\) tends to infinity.
The asymptotic behavior of the integral can be obtained by expanding the integrand around its maximum, and generally, the more terms included in the expansion, the more accurate the asymptotic approximation will be.

In Section \ref{sec:structure}, we have demonstrated that SREE satisfies equipartition up to the constant order \(\mathcal{O}(V^0)\). In that proof, for the case in which \(F'(\mu)=0\) at the maximum point, the integral for the symmetry-resolved partition function needs to be expanded to the subleading order, given by
\begin{equation}
  I(V) \sim C \frac{\mathrm{e}^{V g(c)}}{\sqrt{V}} \left(1 + \frac{a}{V}\right), \, V\rightarrow \infty,
\end{equation}
where \(C\) and \(a\) are constants independent of \(V\).
To achieve this level of approximation, the integrand needs to be expanded as follows:
\begin{multline}
  I(V) \sim \int_{c-\epsilon}^{c+\epsilon}\left[h(c)+h'(c)(\mu-c) + \frac{1}{2}h''(c)(\mu-c)^2\right] \\
  \times \exp\left\{V\left[g(c) + \frac{1}{2}(\mu-c)^2 g''(c) + \frac{1}{6}(\mu-c)^3 g'''(c) \right.\right. + \left.\left.\frac{1}{24}(\mu-c)^4 g^{(4)}(c)\right]\right\} \dd\mu,
\end{multline}
where \(\epsilon\) is a small parameter, indicating that the main contribution to the integral comes from the neighborhood around the maximum \(c\). It is worth noting, however, that the final result of the asymptotic expansion is independent of \(\epsilon\).

In the above expression, we expand the terms in the exponential function beyond the second derivative as follows:
\begin{multline}
  \exp \left\{V\left[\frac{1}{6}(\mu-c)^3 g'''(c) + \frac{1}{24}(\mu-c)^4 g^{(4)}(c)\right]\right\}   \\
  \approx 1+ V\left[\frac{1}{6}(\mu-c)^3 g'''(c) + \frac{1}{24}(\mu-c)^4 g^{(4)}(c)\right] + \frac{1}{72} V^2 (\mu-c)^6 [g'''(c)]^2.
\end{multline}
Substituting this expansion into the previous expression and extending the integration limits to infinity\footnote{It can be ensured that, even though the integration limit is drastically changed from a neighborhood of \(c\) to the whole real line, the resulting asymptotic expansion is valid, basically because the main contribution to the integral comes only from the neighborhood around the maximum.}, we obtain the following integrals:
\begin{multline}
  I(V) \sim \int_{-\infty}^{+\infty} \mathrm{e}^{Vg(c)+V(\mu-c)^2 g''(c)/2}   \\
  \times  \left\{h(c)+\frac{(\mu-c)^2}{2}h''(c) + (\mu-c)^4 \left[\frac{1}{24} V h(c) g^{(4)}(c) + \frac{1}{6}V h'(c) g'''(c)\right]\right.   \\
  \left. + \frac{1}{72}(\mu-c)^6 V^2 h(c) [g'''(c)]^2\right\}\dd\mu, \quad V \rightarrow +\infty.
\end{multline}
Each integral in this expression can be evaluated exactly, yielding the expansion needed for the proof presented in the main text, as given by Equation \eqref{eq:SRpart_2nd_order}:
\begin{equation}
  I(V) = \ee^{V g} \frac{\sqrt{2\pi}}{\sqrt{V (-g'')}} \left(h +
  \frac{-\frac{h''}{2 g''}+\frac{g^{(3)} h'}{2 (g'')^2}-\frac{5 h \left(g^{(3)}\right)^2}{24 (g'')^3}+\frac{h g^{(4)}}{8 (g'')^2}}{V}\right).
\end{equation}
Here, \(g^{(k)}\) denotes the \(k\)-th derivative with respect to \(\mu\), and both the function \(h\) and the function \(g\), along with their derivatives, are evaluated at \(\mu = c\). The appearance of \(-g''(c)\) in the denominator under the square root arises because \(c\) is the point where \(g\) attains its maximum, meaning its second derivative is negative, \(g''(c) < 0\).

\bibliographystyle{JHEP}
\bibliography{ref}

\end{document}